\documentclass[lettersize,journal]{IEEEtran}
\usepackage{amsmath,amsfonts}
\usepackage{algorithmic}
\usepackage{algorithm}
\usepackage{array}
\usepackage[caption=false,font=normalsize,labelfont=sf,textfont=sf]{subfig}
\usepackage{textcomp}
\usepackage{stfloats}
\usepackage{url}
\usepackage{verbatim}
\usepackage{graphicx}
\usepackage{cite}
\usepackage{siunitx}                %
\usepackage{subfig}
\usepackage[export]{adjustbox}
\usepackage{makecell}
\usepackage{scrextend}
\usepackage{booktabs}
\usepackage{bm}		%
\usepackage{hyperref}

\usepackage{tikz}
\usepackage{pgfplots}
\definecolor{color5}{rgb}{0.12156862745098,0.466666666666667,0.705882352941177}	%
\definecolor{color0}{rgb}{0,0,0}	%
\definecolor{color1}{rgb}{1,0.498039215686275,0.0549019607843137}
\definecolor{color2}{rgb}{0.172549019607843,0.627450980392157,0.172549019607843}
\definecolor{color3}{rgb}{0.83921568627451,0.152941176470588,0.156862745098039}
\definecolor{color4}{rgb}{0.580392156862745,0.403921568627451,0.741176470588235}
\definecolor{color6}{rgb}{0.890196078431372,0.466666666666667,0.76078431372549}
\definecolor{color7}{rgb}{0.737254901960784,0.741176470588235,0.133333333333333}
\definecolor{color8}{rgb}{0.549019607843137,0.337254901960784,0.294117647058824}	%

\usepackage{amsmath}
\DeclareMathOperator*{\argmin}{arg\,min}

\newcommand{\PLH}{{\mkern-2mu\times\mkern-2mu}}		%
\newcommand{\commentOut}[1]{}

\usepackage{rotating}		%

\usepackage{tablefootnote}
\usepackage{mymacros}

\newcommand\copyrighttext{
	\footnotesize \textcopyright 2022 IEEE. Personal use of this material is permitted.
	Permission from IEEE must be obtained for all other uses, in any current or future 
	media, including reprinting/republishing this material for advertising or promotional 
	purposes, creating new collective works, for resale or redistribution to servers or
	lists, or reuse of any copyrighted component of this work in other works. 
	DOI: \href{https://doi.org/10.1109/TCSVT.2022.3195322}{10.1109/TCSVT.2022.3195322} }
\newcommand\copyrightnoticeOwn{%
	\begin{tikzpicture}[remember picture,overlay]
		\node[anchor=north,yshift=-10pt, fill=white] at (current page.north) {\fbox{\parbox{\dimexpr\textwidth-\fboxsep-\fboxrule\relax}{\copyrighttext}}};
	\end{tikzpicture}%
}

\begin{document}

\title{Boosting Neural Image Compression for Machines Using Latent Space Masking}

\author{Kristian Fischer,~\IEEEmembership{Student Member,~IEEE,}
Fabian Brand,~\IEEEmembership{Student Member,~IEEE,}	 
Andr\'e Kaup,~\IEEEmembership{Fellow,~IEEE}
\thanks{The authors are with the Chair of Multimedia Communications and Signal Processing Friedrich-Alexander-Universit\"at Erlangen-N\"urnberg, 91058
	Erlangen, Germany (e-mail: kristian.fischer@fau.de; fabian.brand@fau.de; andre.kaup@fau.de)}%
\thanks{Accepted for T-CSVT special issue ``Learned Visual Data Compression for both Human and Machine'' on 21st of Jul. 2022; DOI: \href{https://doi.org/10.1109/TCSVT.2022.3195322}{10.1109/TCSVT.2022.3195322}}}

\maketitle
\copyrightnoticeOwn

\begin{abstract}
	Today, many image coding scenarios do not have a human as final intended user, but rather a machine fulfilling computer vision tasks on the decoded image. Thereby, the primary goal is not to keep visual quality but maintain the task accuracy of the machine for a given bitrate. Due to the tremendous progress of deep neural networks setting benchmarking results, mostly neural networks are employed to solve the analysis tasks at the decoder side. Moreover, neural networks have also found their way into the field of image compression recently. These two developments allow for an end-to-end training of the neural compression network for an analysis network as information sink.
	Therefore, we first roll out such a training with a task-specific loss to enhance the coding performance of neural compression networks. Compared to the standard VVC, 41.4\,\% of bitrate are saved by this method for Mask R-CNN as analysis network on the uncompressed Cityscapes dataset. As a main contribution, we propose LSMnet, a network that runs in parallel to the encoder network and masks out elements of the latent space that are presumably not required for the analysis network. By this approach, additional 27.3\,\% of bitrate are saved compared to the basic neural compression network optimized with the task loss. 
	In addition, we are the first to utilize a feature-based distortion in the training loss within the context of machine-to-machine communication, which allows for a training without annotated data. We provide extensive analyses on the Cityscapes dataset including cross-evaluation with different analysis networks and present exemplary visual results.
	
	Inference code and pre-trained models are published at \\ \url{https://github.com/FAU-LMS/NCN_for_M2M}.
	%% WE propose LSM net
\end{abstract}


\section{Introduction}

Standard image and video coding algorithms such as JPEG~\cite{wallace1992_JPEG}, High Efficiency Video Coding~(HEVC)~\cite{sullivan2012_HEVC}, and Versatile Video Coding~(VVC)~\cite{chen2020vtm10} are commonly optimized for the human visual system~(HVS).
% Add VP9 here?
However, more and more of today's Internet traffic is accounted to machine-to-machine~(M2M) communication, where the transmitted visual data is analyzed by computer algorithms fulfilling a certain task. Cisco~\cite{cisco2020} predicts that half of all devices and connections will be accounted to M2M communication by 2023. This leads to a major challenge of transmitting and storing all the incoming data demanding for novel coding schemes that are specifically optimized for machines and algorithms as information sink. Thus, MPEG introduced an ad-hoc group on \textit{Video Coding for Machines~(VCM)} in 2019~\cite{zhang2019} with the ultimate goal of standardizing a bitstream format that is optimal for M2M scenarios.

An exemplary field of application for VCM is intelligent transportation. There, multiple cars with their installed sensors can group together as a visual sensor network~\cite{soro2009,tavli2012}. The captured data can be transmitted with low latency to nearby edge nodes~\cite{satyanarayanan2017, liu2019}, where computationally expensive algorithms are applied to the data in order to derive important information such as the detection of other road participants. With that, possible hazards can be detected and transfered back to the cars to initiate countermeasures. The major data share can be accounted to the visual data captured by the cameras installed in the cars. According to Intel, 4\,TB of data are generated by one car within 1.5 hours of driving~\cite{winter2017}. Despite the increasing presence of 5G networks, it is infeasible to share all that information with the edge nodes without optimal compression particularly at places with a high traffic volume. Similar challenges with numerous sensor nodes transferring multimedia data to edge computing platforms also rise for other Internet of things fields such as surveillance of public spaces~\cite{mosaif2021ANS} or Industry 4.0~\cite{narayanan2020}.

\begin{figure}[!t]
	\centering
	\includegraphics[width=\linewidth]{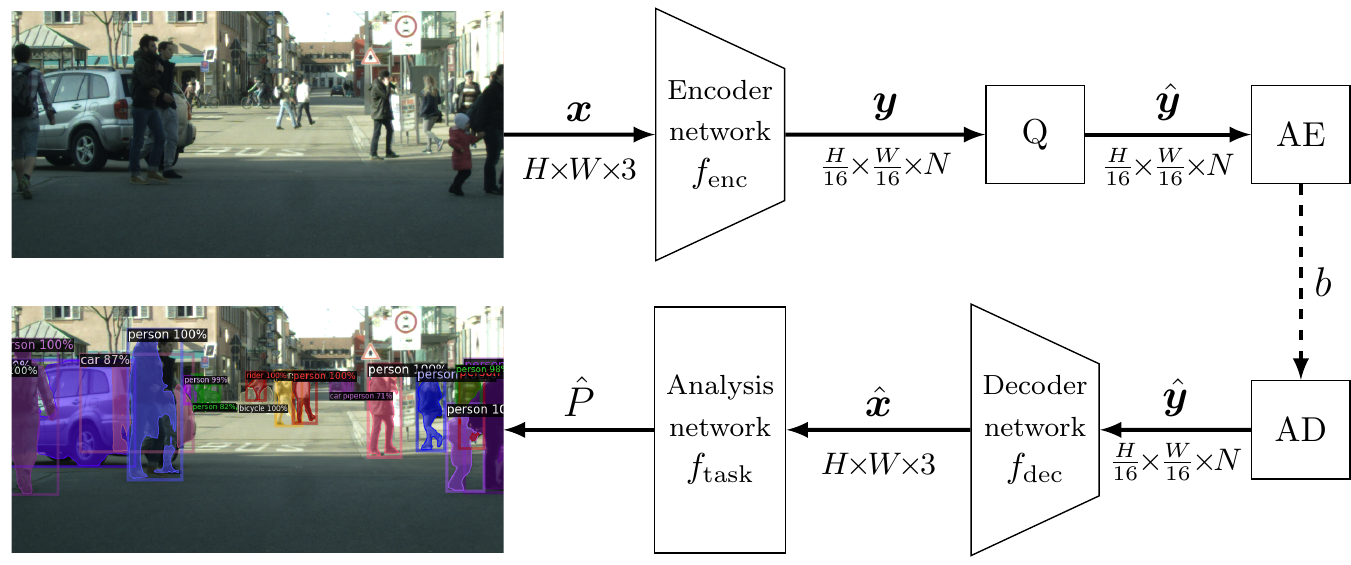}
	\caption{Proposed neural compression framework for M2M communication with instance segmentation as analysis task. The upper and the lower branch symbolize the encoder and the decoder side respectively. The hyperprior has been omitted for simplicity. Below each signal, its dimensionality is given in the order \textit{height} $\!\times\!$ \textit{width} $\!\times\!$ \textit{number of channels}. $Q=$ quantization; $AE=$~arithmetic encoder; $AD=$ arithmetic decoder.}
	\label{fig:proposed compression framework}
	\vspace{-5mm}
\end{figure}

In those depicted M2M scenarios, the key challenge is to reduce the bitrate of the data transfered to the edge. By that, a similar performance of the algorithms applied to the decoded, deteriorated input data solving different computer vision tasks such as object detection, semantic segmentation, or tracking has to be maintained. In the recent past, mostly neural-network-based algorithms set the benchmark in solving such problems. Also most prior VCM approaches are partially based on machine learning
%to improve the VCM-performance significantly over standard hybrid codecs
~\cite{galteri2018, choi2018 ,fischer2021_ICASSP, fischer2020_FRDO}.

In this paper, we aim at boosting the image coding performance for machines by employing fully neural-network-based codecs, as presented amongst others in~\cite{balle2017endtoend, balle2018_bmshj2018, minnen2018_mbt2018, chen2021}. With that, the trade-off between a high compression and a high performance of the applied analysis network is optimized in an end-to-end fashion. To improve the standard neural compression network~(NCN) architectures similar to~\cite{minnen2018_mbt2018} for VCM scenarios, we propose the novel additional latent space masking network (LSMnet), which masks out non-salient components of the latent space in order to significantly reduce the bitrate but still maintain the detection accuracy. Moreover, we propose to employ a new training loss based on a feature-based distortion inspired by~\cite{fischer2020_FRDO} that allows to adapt NCNs for the VCM task without requiring annotated training data.

The general proposed NCN-based VCM framework is depicted in Figure~\ref{fig:proposed compression framework}. The input RGB image $\inputImage$ with the height $\height$ and the width $\width$ is fed into the neural encoder network $\encoderNetwork$ returning the latent representation $\latentRep$ of lower spatial resolution and $\numChannels$ channels. $\latentRep$ is then quantized into $\latentRepQuant$ before being fed into a lossless arithmetic coder. At the decoder side, the decoder network $\decoderNetwork$ generates the deteriorated output image $\outputImage$ from the transmitted $\latentRepQuant$. For standard coding for the HVS, the distortion between $\inputImage$ and $\outputImage$ could now be calculated. For VCM, $\outputImage$ is fed into the analysis network performing its task resulting in the predictions $\predictionsHat$. Ultimately, the predictions are compared against the ground-truth data in order to measure the performance of the analysis network. The compression rate is derived from the length of the bitstream~$b$.

NCNs are typically based on an autoencoder structure~\cite{balle2017endtoend} and are trained to reduce the following rate-distortion loss, similar to the rate-distortion optimization in hybrid codecs
\begin{equation}
	\lossNcn = \distortion(\inputImage,\outputImage) + \lambda \cdot \rate(\latentRepQuant).
	\label{eq:training loss}
\end{equation}
The trade-off between a low rate $\rate$ to transmit the latent $\latentRepQuant$ and a low distortion $\distortion$ can be steered via the user-defined parameter $\lambda$ during training. When optimizing NCNs for the HVS, $\distortion$ is typically measured by the mean squared error~(MSE) or the structural similarity index~(SSIM\footnote{Throughout this paper, we employed MS-SSIM. For simplicity, we will denote this as SSIM in the following.})~\cite{wang2004}. However, previous research~\cite{fischer2021_ICIP} revealed that improving coding schemes for the HVS does not necessarily result in a higher coding efficiency for machines. Thus, the key target of this paper is to enhance the coding performance for NCNs in VCM scenarios. All in all, our paper provides the following contributions:
\begin{itemize}
	\item First, we employ an NCN with additional hyperprior similar to~\cite{minnen2018_mbt2018} and adapt its training loss from \eqref{eq:training loss} for the VCM task of instance segmentation resulting in an \textbf{end-to-end optimized NCN model} for VCM scenarios. Therefore, we consider the loss of the analysis network as task-specific distortion replacing the HVS distortion similar to~\cite{le2021_ICASSP, chamain2021}. 	
	Contrary to~\cite{chamain2021}, we evaluate our results on uncompressed input data which lowers the possibility of undesired effects induced by already existing coding artifacts.
	Besides, contrary to~\cite{le2021_ICASSP}, we applied analysis models that have only been trained on uncompressed data as well, in order to avoid effects of an analysis model with an increased robustness when being applied on compressed data as shown by~\cite{poyser2020,fischer2021_ISCAS}. 
	\item Moreover, we employ a \textbf{feature-based distortion loss} for the first time in the VCM context that improves the coding performance by a large margin.
	There, the distortion is measured in the features generated by the backbone of the analysis network.
	The major advantage that comes with this method is that the training does not require labeled ground truth data, which is very cumbersome and costly to get in practical scenarios. %% Maybe also better for cross-evaluation
	\item As our main contribution, we propose the novel masking network \textbf{LSMnet} that runs in parallel to the NCN encoder and masks out certain values in the latent space $\latentRep$ resulting in less bits to transmit. Thereby, LSMnet is derived from the early features created by the analysis network allowing for a targeted masking that greatly reduces the bitrate but does not harm the performance of the analysis network.
	\item Finally, we provide extensive studies for the tasks of object detection and segmentation proving that the proposed NCN-based VCM framework and LSMnet outperform standard approaches optimized for the HVS. In addition, cross evaluations are shown where the analysis models used in training and inference differ.
\end{itemize}

The remainder of this paper is organized as follows: The next chapter summaries related work in terms of VCM and NCNs. Subsequently, a detailed overview on the components of our VCM coding framework scenario from Figure~\ref{fig:proposed compression framework} is given. Thereupon, the proposed adapted loss functions for the VCM-optimized NCNs are presented before introducing the novel additional LSMnet to mask the latent space. Afterwards, the evaluation results are shown and conclusions will be given.

\section{Related Work}

%The following chapter summarizes the most important literature in the fields of VCM and NCNs.

\subsection{Image/Video Coding for Machines}

The compression of visual data for M2M communication can follow two different principles~\cite{redondi2013, duan2020_IEEE_TIP}: \textit{Analyze-then-compress} implies that the main computation is directly executed at each sensor, and that the derived features are transferred to the edge nodes resulting in high bitrate savings. When following the \textit{compress-then-analyze} paradigm, the sensor data is first transmitted to a remote processing server, where the features are derived. This procedure is commonly more energy efficient and allows for low-cost sensors. Besides, the data is still interpretable for e.g. a human supervisor which allows to comprehend the drawn decisions of the machine. For the scope of this paper, we will focus on the latter paradigm.

The negative influence of artifacts and information loss induced by compression on applied algorithms has been discussed in various papers. Bokar et al.~\cite{borkar2019_IEEE} showed the performance gap when the neural network is trained on pristine data, but the inference runs on differently distorted data\commentOut{and propose methods to correct that gap}. In an earlier work~\cite{fischer2020_ICIP}, we demonstrate that the coding gains of VVC over its predecessor HEVC are significantly lower when coding for instance detection and segmentation networks instead of the HVS. In~\cite{fischer2021_ISCAS}, we proposed to add compressed image data into the training procedure of an instance segmentation network to greatly enhance its robustness against VVC compression artifacts during inference. Poyser et al.~\cite{poyser2020} measured the negative influence of JPEG and H.264 coding on the performance of deep neural networks solving five different neural network tasks.

Several optimizations have been proposed to improve the coding performance of standard hybrid codecs. Before the breakthrough of deep neural networks, coding for machines has mostly been referred to as feature-preserving coding. Duan et al.~\cite{duan2012} and Chao et al.~\cite{chao2013} proposed an adapted quantization table to scale the discrete cosine transform~(DCT) coefficients during JPEG compression such that the performance of visual search algorithms and scale-space-based detectors was improved for limited bitrate. Throughout the last years, deep neural networks have set benchmarks in most fields of computer vision. Thus, they are also mostly used as information sink for evaluating M2M coding. Choi et al.~\cite{choi2018} proposed a rate-control algorithm for HEVC that assigns the bits depending on an importance map derived from the early layers of the analysis network for object detection. Similar approaches with saliency coding were done in~\cite{galteri2018} and \cite{fischer2021_ICASSP} for HEVC and VVC, respectively. Furthermore, we proposed an enhanced rate-distortion optimization for VCM coding in~\cite{fischer2020_FRDO}, which measures the distortion inside VVC in the feature space of an image classification network rather than in the pixel domain resulting in bitrate savings over standard VVC. 

Applying NCN architectures to perform an end-to-end training of the compression network and the analysis network to upgrade the coding performance for VCM scenarios is a very recent idea. In a previous paper~\cite{fischer2021_ICIP} we evaluated different NCN architectures trained for the HVS on their VCM performance for instance segmentation. There, it was shown that training on SSIM provides higher coding gains on VCM than MSE, and that the GAN-based NCN HiFiC~\cite{mentzer2020hific} is even able to outperform VVC on the Cityscapes~\cite{cordts2016} dataset. Le et al.~\cite{le2021_ICASSP} and Chamain et al.~\cite{chamain2021} applied an end-to-end-training of NCNs with a task specific loss on instance segmentation and detection resulting in large coding gains over standard VVC. 

%But, these works have the drawback that they are both trained and evaluated on the COCO dataset~\cite{lin2014_COCO} whose images already contain major deteriorations induced by JPEG coding and it is not clear which influence those have on the coding performance. In contrast, our models in this paper are trained and evaluated on the pristine and uncompressed Cityscapes dataset eliminating the possibility of undesired effects from predominant coding deteriorations in the training and inference data. 

% Possible papers to add: Le et al. and Patwa et al.

%\begin{itemize}
%	\item Feature-preserving image/video coding
%	\item Coding for neural networks with standard hybrid image/video codecs
%	\item Coding with neural compression networks for machines \cite{fischer2021_ICIP, le2021_ICASSP, wang2021, le2021_contentAdaptive}
%\end{itemize}

\subsection{Neural Compression Networks}
%In standard hybrid codecs, orthogonal transformations, e.g. DCT or wavelet functions, shift the signal energy on a small number of coefficients. Subsequently, the coefficients are quantized and entropy coded. Highly non-linear methods such as prediction and block partitioning are further employed to raise the coding performance for the non-linear input data.

With the increased knowledge and applicability of neural networks, end-to-end-learned neural image compression has risen throughout the last five years.
%One of the earliest work in this field~\cite{toderici2017} applied a coding chain of encoder, binarizer, and decoder implemented as recurrent neural network.
The pioneering work in this field was published by Ball\' e et al.~\cite{balle2017endtoend}, where he proposed the idea to learn a non-linear convolutional neural network~(CNN) as encoder network generating a compact latent representation with reduced dimensionality. After quantization, the representation is entropy coded in a lossless manner. The complementary decoder network reconstructs the input image from the transmitted latent representation. Replacing the quantization step by adding uniform noise during training allows for an end-to-end optimization of the variational autoencoder-based~\cite{kingma2014} structure. Besides, this work deployed the generalized divisive normalization~(GDN) transformation~\cite{balle2016density} as non-linearity inspired by visual systems in nature and performs well in modeling natural images.

Follow-up work by Ball\' e et al.~\cite{balle2018_bmshj2018} introduced an additional hyperprior, employed to model the probability distributions of the latent space for the entropy coding by zero-mean Gaussians. Minnen et al.~\cite{minnen2018_mbt2018} improved the autoencoder performance by also estimating the mean and introduced an autoregressive context model in order to further reduce the entropy of the latent representation. Their NCN was the first to outperform the HEVC-based image codec on SSIM and PSNR. The work in~\cite{cheng2020} claimed to be the first NCN performing on par with VVC intra coding in terms of PSNR by employing attention models to focus more on complex regions. More recent work in~\cite{guo2021} additionally exploits cross-channel redundancies in the latent space and utilizes global reference points for a more compact latent representation. \commentOut{resulting in coding gains over VVC intra coding.} \commentOut{Chen et al.~\cite{chen2021} employed global attention masks for NCNs.}

%\begin{itemize}
%	\item Standard chain from Ball\'e
%	\item More recent approaches that are also close to VVC aiming on further minimizing the redundancy in the latent space by improving the context model
%	\item maybe 2 examples for video coding
%\end{itemize}

\section{Coding Framework for VCM Scenario}

Throughout this paper, we optimize NCNs for the inference case that they are used in a VCM scenario\commentOut{ where they code a single image $\inputImage$ for an arbitrary analysis network applied to the compressed output image $\outputImage$} as depicted in Figure~\ref{fig:proposed compression framework}. In general, arbitrary combinations of NCN and analysis network can be employed for our proposed approach. Naturally, the method results in the highest coding gains, when the combination remains the same for training and inference. For the scope of this paper, we select a hyperprior-based autoencoder similar to Minen et al.~\cite{minnen2018_mbt2018} as NCN since it has become the de facto standard in the NCN community. Besides, the network structure is simple and allows for fast training of the proposed methods. As analysis network to optimize for throughout the training process, we employ the state-of-the-art instance segmentation network Mask R-CNN~\cite{he2017} with a ResNet-50~\cite{he2016resnet} backbone with feature pyramid structure~(FPN)~\cite{lin2017fpn}. 

%\begin{itemize}
%	\item Explain coding framework with the help of Figure 1
%	\item Define that we are only dealing with images and mainly focus on instance segmentation with Mask R-CNN and ResNet FPN 50
%	\item But, training procedure can be applied for other tasks and networks as well.
%\end{itemize}

\begin{figure*}[!t]
	\centering
	\includegraphics[width=0.95\linewidth]{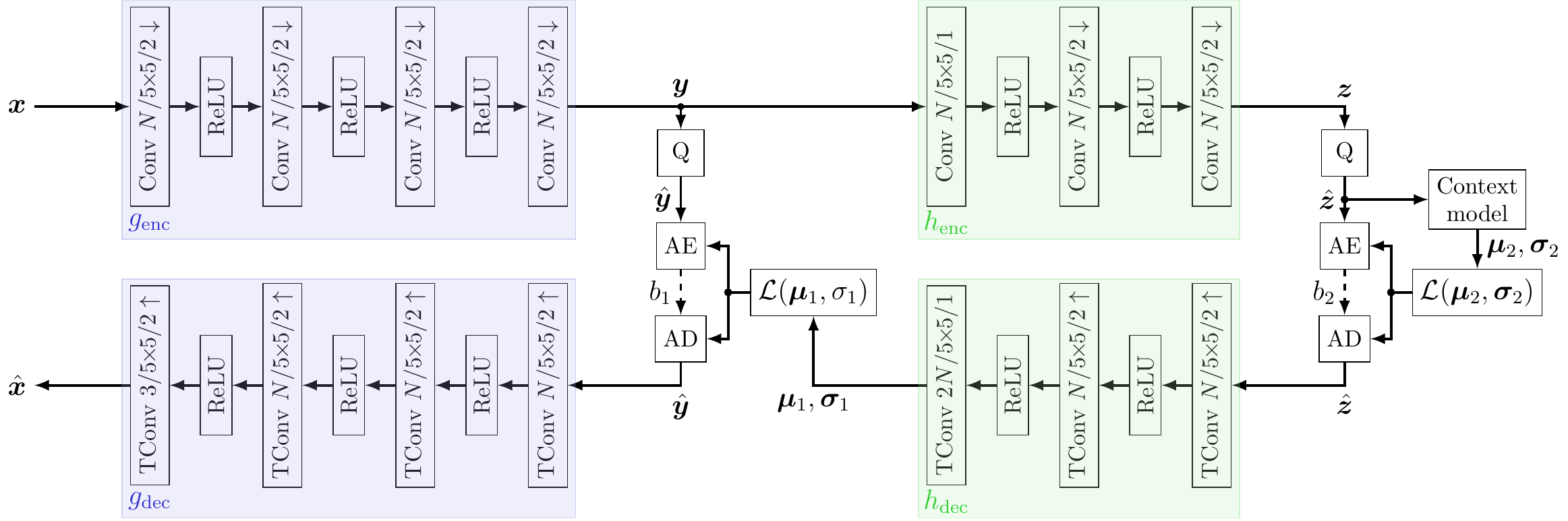}
	\caption{Network structure of employed NCN combining core autoencoder (blue) and hyperprior coder (green) similar to~\cite{minnen2018_mbt2018}. \textit{Conv} $c/k\PLH k/s\downarrow$ and \textit{TConv} $c/k\PLH k/s\uparrow$ denote a convolutional and transposed convolutional layer, respectively, with the number of channels $c$, the kernel size $k$, and the stride $s$. $\mathcal{L}$ denotes the Laplace distribution with mean $\mu$ and standard variation $\sigma$. $b_1$ and $b_2$ denote the bitstreams required to transmit the latent representation $\latentRepQuant$ and the hyperprior $\hyperpriorQuant$, respectively.}
	\label{fig:used NCN structure}
\end{figure*}

\subsection{Basic Neural Compression Network}
\label{subsec: Basic Neural Compression Network}
\subsubsection{Structure}

The basic structure of the used NCN is similar to~\cite{minnen2018_mbt2018} and is shown in Figure~\ref{fig:used NCN structure}. For the core autoencoder, the input image $\inputImage$ is fed into the encoder network $\encoderCoreNcn$, which compresses the image information into a latent space $\latentRep$ by convolutional layers with a stride of two and subsequent non-linearity.
% Comparison between the different non-linearities?
Afterwards, $\latentRep$ is quantized and the resulting $\latentRepQuant$ is losslessly transmitted in bitstream $b_1$ to the receiver side. There, the decoder network $g_\mathrm{dec}$ reconstructs the compressed image $\outputImage$ from the transmitted $\latentRepQuant$. 
%Thereby, the compression of such NCNs is rather related to a more compact representation of the image information in the latent space $\latentRep$ that is more efficient to transmit than the reduction of spatial dimensionality.

In order to exploit the spatial dependencies, which are still present in $\latentRepQuant$, for a more efficient transmission, an additional hyperprior network is attached to the core autoencoder which estimates the mean $\mean_1$ and the variance $\stdDeviation_1$ of a Laplacian distribution $\laplaceFunc$ to model the probability densities for each element of $\latentRepQuant$. By providing suitable models, the entropy of $\latentRepQuant$, and thus the bitrate, can be significantly reduced. 
We selected Laplacian distributions over Gaussian models used in~\cite{minnen2018_mbt2018} since this was also shown to perform better in~\cite{zhou2018VariationalAF} and also our preliminary experiments showed a superior performance of using Laplacian models.
The hyperprior network is also build up as an autoencoder. First, the latent representation $\latentRep$ is fed into the hyper encoder $\encoderHypNcn$ generating the hyperprior latent space $\hyperprior$. Similar to the core autoencoder, $\hyperprior$ is quantized into $\hyperpriorQuant$ and transmitted via a second bitstream $\bitstream_2$. At the decoder side, $\mean_1$ and $\stdDeviation_1$ are generated from $\hyperpriorQuant$ by the hyper decoder $\decoderHypNcn$. An autoregressive context model is applied to $\hyperpriorQuant$ modeling a second Laplacian distribution to transmit the hyperprior $\hyperpriorQuant$ as efficient as possible. This differs from the reference approach in~\cite{minnen2018_mbt2018}, where the autoregressive model is applied in addition to the hyperprior to generate the first Laplacian distribution.

Another difference to~\cite{minnen2018_mbt2018} is that we selected rectified linear units~(ReLUs) as non-linearity over GDNs~\cite{balle2016density}, since they are the main non-linearity in the ResNet backbone of the analysis network to optimize for. In our previous research\cite{fischer2021_ICIP}, we showed that NCNs with non-linearities from the ReLU family perform better than networks with GDN as non-linearity in VCM scenarios. 

\subsubsection{Training}
The learnable network parameters~$\parameters$ of the four sub-networks $\encoderCoreNcn$, $\decoderCoreNcn$, $\encoderHypNcn$, and $\decoderHypNcn$ are jointly trained in an end-to-end manner. 
%The given NCN structure allows for a joint end-to-end training of the four sub networks $\encoderCoreNcn$, $\decoderCoreNcn$, $\encoderHypNcn$, and $\decoderHypNcn$ that all depend on the learnable network parameters~$\parameters$. 
Since the hyperprior is transmitted in a separate bitstream, the bitrate in the training loss given in~\eqref{eq:training loss} depends on both latents $\latentRepQuant$ and $\hyperpriorQuant$. 
%\begin{equation}
%	\lossNcn = \distortion(\inputImage,\outputImage) + \lambda \cdot (\rate(\latentRepQuant) + \rate(\hyperpriorQuant)).
%	\label{eq:loss function 2}
%\end{equation}
In general, the optimal network parameters are derived by simultaneously minimizing the estimated rate to transmit the latent representations and reducing the distortion between $\inputImage$ and $\outputImage$:
\begin{equation}
	\begin{aligned}
		%	\parameters = \argmin_{\parameters} \distortion(\inputImage, \decoderCoreNcn(\latentRepQuant|\parameters)) + \\ \lambda \cdot (\rate(\encoderCoreNcn(\inputImage|\decoderHypNcn(\hyperpriorQuant|\parameters),\parameters)) + \rate(\encoderHypNcn(\latentRep|\parameters))).
		% Simplified due to reviewers comment
		\parameters = \argmin_{\parameters} \distortion(\inputImage, \decoderCoreNcn(\latentRepQuant|\parameters)) + \\ \lambda \cdot (\rate(\encoderCoreNcn(\inputImage|\parameters)) + \rate(\encoderHypNcn(\latentRep|\parameters))).
	\end{aligned}
	\label{eq:loss NCN general}
\end{equation}
As distortion for the regular use case when optimizing the network for the HVS, we employed the same combination of SSIM and MSE as distortion similar to~\cite{brand2021_CVPR}
\begin{equation}
	\distortionHvs(\inputImage, \outputImage) = \distortion_\mathrm{MSE}(\inputImage, \outputImage) + 0.1\cdot D_\mathrm{SSIM}(\inputImage, \outputImage).
	\label{eq: distortion function HVS}
\end{equation}
Eventually, this results in the following HVS-based loss for the complete NCN $\ncnComplete$:
\begin{equation}
	\lossHvs = \distortionHvs(\inputImage,\ncnComplete(\inputImage|\parameters)) + \lambda \cdot \rate(\ncnComplete(\inputImage|\parameters)).
	\label{eq: loss function HVS}
\end{equation}
The network optimized for this HVS-based loss serves as reference NCN for our research in the remainder of this paper.

%\begin{figure}[!t]
%	\centering
%	\includegraphics[width=0.9\linewidth]{tikz_mask_rcnn/tikz_mask_rcnn.pdf}
%	\vspace{-3mm}
%	\caption{Mask R-CNN top-level structure.}
%	\label{fig:structure mask R-CNN}
%\end{figure}

%\begin{itemize}
%	\item For the scope of our paper we select a basic NCN with additional hyperprior~\cite{minnen2018_mbt2018} in order to keep the complexity reasonable
%	\item Structure as the basemodel in~\cite{brand2021_CVPR}
%	\item Figure with structure and Mask R-CNN in the end (probably over two columns)
%\end{itemize}

\subsection{Mask R-CNN as Evaluation Network}
\subsubsection{High-Level Structure}
The main goal of this paper is to optimize and adapt the previously presented NCN for neural networks as information sink. To that end, we selected the task of instance segmentation with Mask R-CNN to primarily optimize for. In Mask R-CNN, the input image is first fed into the backbone with a CNN structure to obtain high-level semantic feature maps. From those features, the region proposal network~(RPN) estimates region of interest~(RoI) proposals. Thereupon, each feature map that corresponds to a RoI is cut out and fed into the RoI pooling layer which returns a feature map of fixed spatial size for each RoI. Ultimately, the RoI head is applied to the feature vector to derive a prediction consisting of a class label with a certainty score, a bounding box, and a binary mask for each RoI.

During Mask R-CNN training, five different losses from the RPN and the RoI head network are jointly minimized
\begin{equation}
	\lossMaskRcnn = L_\mathrm{RPN\_cls} + L_\mathrm{RPN\_loc} + L_\mathrm{cls} + L_\mathrm{box} + L_\mathrm{mask}.
	\label{eq: mask R-CNN loss}
\end{equation}
$L_\mathrm{RPN\_cls}$ and $L_\mathrm{RPN\_loc}$ are the classification and localization loss of the RPN layer, respectively. $L_\mathrm{cls}$ defines how well the RoI head preforms in predicting the class labels. $L_\mathrm{box}$ and $L_\mathrm{mask}$ define the accuracy of the predicted bounding boxes and binary masks, respectively.

\subsubsection{ResNet-50 FPN Backbone}

\begin{figure}[!t]
	\centering
	\includegraphics[width=0.8\linewidth]{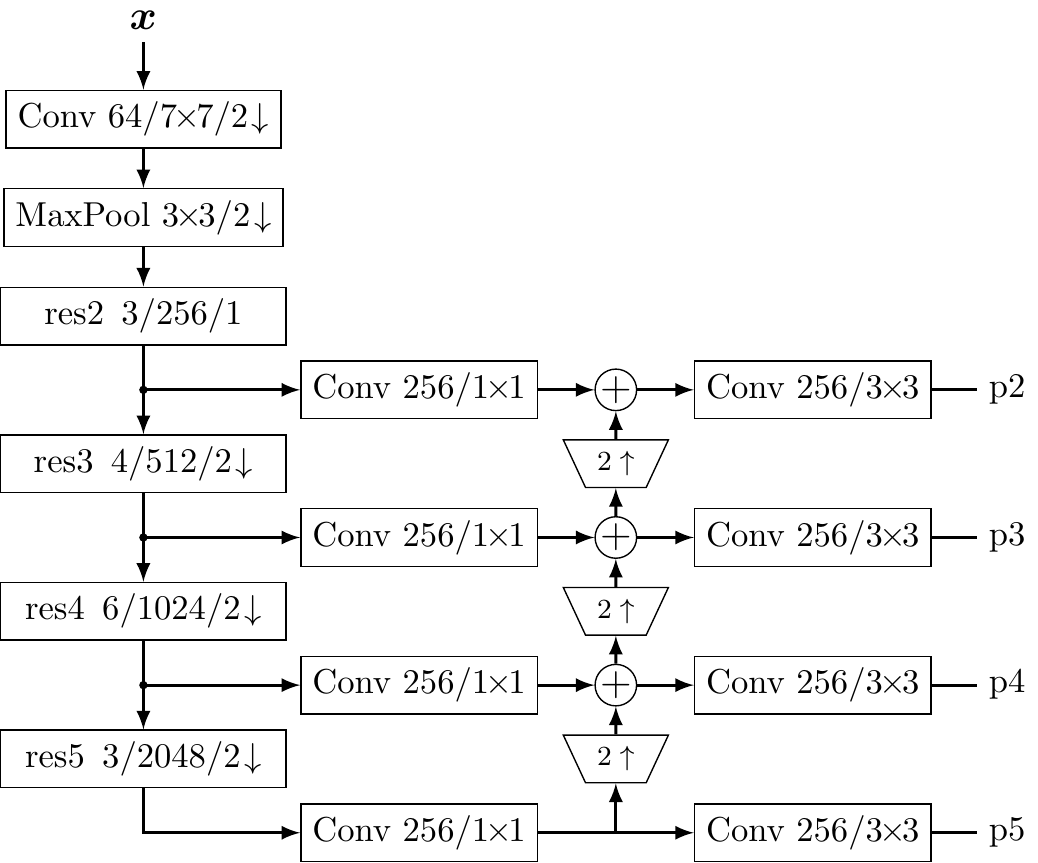}
	\caption{ResNet-50 architecture with FPN structure that is employed as Mask R-CNN backbone with the bottom-up branch on the left.	
		\textit{Conv} $c/k\PLH k$ denotes a convolution layer with $c$ output channels and a $k\PLH k$ kernel.
		The initial convolution layer has a stride of~2. Since all lateral convolutions have a stride of 1, it is omitted in the drawing.
		\textit{MaxPool} denotes a max-pooling layer with a kernel size of $3\PLH 3$ and a stride of 2.
		\textit{res*} $\beta/c/s\mkern-3mu\downarrow$ denotes a ResNet stage with $\beta$ bottleneck blocks, $c$ output channels, and the stride $s$.
		The $2\mkern-3mu\uparrow$ blocks symbolize a spatial feature upscaling by a factor of 2 using nearest neighbor interpolation.
		The adder symbol represents an element-wise addition of feature maps.
		Feature map \textit{p6} is not included in this drawing and is generated by spatially sub-sampling \textit{p5} with a stride of 2.}
	\label{fig:structure ResNet 50}
\end{figure}

The chosen backbone for the Mask R-CNN employed throughout this paper is a ResNet-50 with FPN structure as shown in Figure~\ref{fig:structure ResNet 50}. Its task is to extract semantically meaningful features $\featureSpace$ at different spatial resolutions from the input image $\inputImage$. To that end, the bottom-up branch applies a ResNet-50 CNN to $\inputImage$. It consists of a stem followed by four stages that are denoted as \textit{res2}, \textit{res3}, \textit{res4}, and \textit{res5}. Each stage consists of multiple stacked bottleneck blocks~(BBs)~\cite{he2016resnet}, where the first BB of each stage reduces the spatial dimensionality by a factor of two except for \textit{res2}. 
%In the end, the deepest feature map with the semantically most meaningful features has a reduced spatial dimensionality by a factor of 64 in each direction compared to the input $\inputImage$, but 2048 channels. 

The purpose of the FPN structure is to transform these meaningful deep features also into earlier, less meaningful features with a higher resolution. Therefore, the top-down pathway with lateral connections is applied by spatially upscaling the low-resolution features and adding them to the semantically weaker features from the previous stage. Subsequently, each resulting feature map is fed into a $3\PLH 3$ convolutional layer to obtain the four final feature maps with 256 channels but of different spatial resolutions, which are called \textit{p2}, \textit{p3}, \textit{p4} and \textit{p5}. The smallest feature map \textit{p6} is derived by spatially subsampling \textit{p5} and a factor of two. Finally, the RPN of Mask R-CNN can be applied to each of those feature maps to predict the RoIs for the to-be-detected objects. 
%Thanks to the FPN structure, meaningful features can be derived for several spatial resolutions resulting in optimal bounding box proposals for differently sized objects.

%\begin{itemize}
%	\item Explain Mask R-CNN structure and ResNet FPN backbone
%	\item Explain Mask R-CNN loss
%	\item Figure with ResNet 50 FPN backbone
%\end{itemize}

\section{VCM-Optimized Feature-Based Loss}

In our previous works~\cite{fischer2020_ICIP, fischer2020_FRDO ,fischer2021_ICIP} we found that optimizing compression schemes on classic distortion metrics such as MSE in the pixel space does not necessarily result in an optimal compression performance for neural networks as information sink.
We consider two possibilities to optimize the NCNs for VCM scenarios. First, we employ already existing schemes which end-to-end train the NCN on the task loss. Second, we propose a novel feature-based loss which does not require hand-annotated data.

\subsection{Task-Driven Loss}
The idea of substituting the commonly employed distortion metrics such as MSE or SSIM with the task loss $\lossTask$ of the analysis network has already been investigated recently in~\cite{le2021_ICASSP} and~\cite{chamain2021}. Thereby, the NCN $\ncnComplete(\inputImage,\parameters)$ with its weights $\parameters$ is trained to generate an output image $\outputImage$ that results in the lowest loss $\lossTask$ of the analysis network $\evalNet$ with respect to the required rate $\rate$. 

When training the analysis network $\evalNet$, the loss $\lossTask(\evalNet(\inputImage|\parametersEvalNet))$ depends on the input image $\inputImage$ and its trainable weights $\parametersEvalNet$. In general, the input images are fixed and uncompressed, and the weights $\parametersEvalNet$ are optimized by minimizing $\lossTask$. When training the NCN for VCM scenarios in end-to-end manner, the task network is considered as discriminator to train the NCN $\ncnComplete(\inputImage|\parameters)$ and its network weights $\parametersEvalNet$ remain fixed. During the training, the task network is fed with the coded images that are generated by the differently trained NCN. 
Its performance on the deteriorated images defines the eventual quality of the coding process.
Therefore, the VCM-optimized weights $\parameters$ are found by
\begin{equation}
	\parameters = \argmin_{\parameters} \lossTask(\evalNet(\ncnComplete(\inputImage|\parameters)|\parametersEvalNet)) + \lambda \cdot \rate (\ncnComplete(\inputImage|\parameters)).
	\label{eq: task loss}
\end{equation}
With that, an end-to-end-trained optimization of NCNs for an analysis network fulfilling an arbitrary computer vision task can be performed.
In the remainder of this paper, we specifically train our NCN model with Mask R-CNN as task network and its loss as introduced in \eqref{eq: mask R-CNN loss} such that $\lossTask=\lossMaskRcnn$.

\subsection{Proposed Feature-Based Loss}
The major drawback of the loss function given in \eqref{eq: task loss} is that it requires annotated ground-truth labels to calculate the task-specific loss $\lossTask$. Such datasets are not always available, especially under the constraint of uncompressed training images. 
%Besides, optimizing an NCN for a specific loss of one analysis network $\evalNet$ might not result in rate savings when coding images for other analysis networks or tasks for inference due to the strong adaptation towards the training analysis network. 

Therefore, we propose a novel loss function for VCM scenarios, which is inspired by the successful feature-based rate-distortion optimization~\cite{fischer2020_FRDO}, to improve the NCN training for VCM scenarios. In~\cite{fischer2020_FRDO}, the distortion of the rate-distortion optimization in the VVC encoder is measured in the feature space rather than in the pixel domain which allows the encoder to omit information that is not as present in later feature maps and therewith not that important for solving the evaluation task. Transfered to the VCM optimization of the NCN, the training of the NCN parameters changes to the following minimization problem:
\begin{equation}
	\parameters = \argmin_{\parameters} \distortionFeature(\inputImage, \ncnComplete(\inputImage| \parameters)) + \lambda \cdot \rate (\ncnComplete(\inputImage| \parameters)).
	\label{eq: FB loss}
\end{equation}
The feature-based distortion $\distortionFeature$ is measured by
\begin{equation}
	\distortionFeature = \sum_{i}^{}(\featureSpace[i] - \hat{\featureSpace}[i])^2,
\end{equation}
calculating the SSE between the feature spaces $\featureSpace=\backboneNet(\inputImage)$ and $\hat{\featureSpace}=\backboneNet(\outputImage)$. Contrary to~\cite{fischer2020_FRDO}, the features $\featureSpace$ are not generated by an auxiliary network to ensure a higher generality, but rather the specific backbone $\backboneNet$ of the analysis network which the NCN shall be optimized for. $i$ represents the element-wise access to each element of the three-dimensional feature space. With that novel feature-based loss $\lossFeature$, the NCN learns to return an output image $\outputImage$ that results in high fidelity feature maps $\hat{\featureSpace}$ and thus, high quality detections of the analysis network.
%The task-driven loss can be seen as a special-case of the feature-based loss, where the very last features of the network are taken to measure the quality for the NCN training.
%The best coding performance for both loss functions is expected when the analysis model remains the same during training and inference.

%Furthermore, additional bitrate can be saved for VCM by reducing the amount of spent bits in non-salient areas as shown in ~\cite{galteri2018, choi2018, fischer2021_ICASSP}. 
\section{Latent Space Masking by LSMnet}
\begin{figure}[!t]
	\centering
	\includegraphics[width=0.9\linewidth]{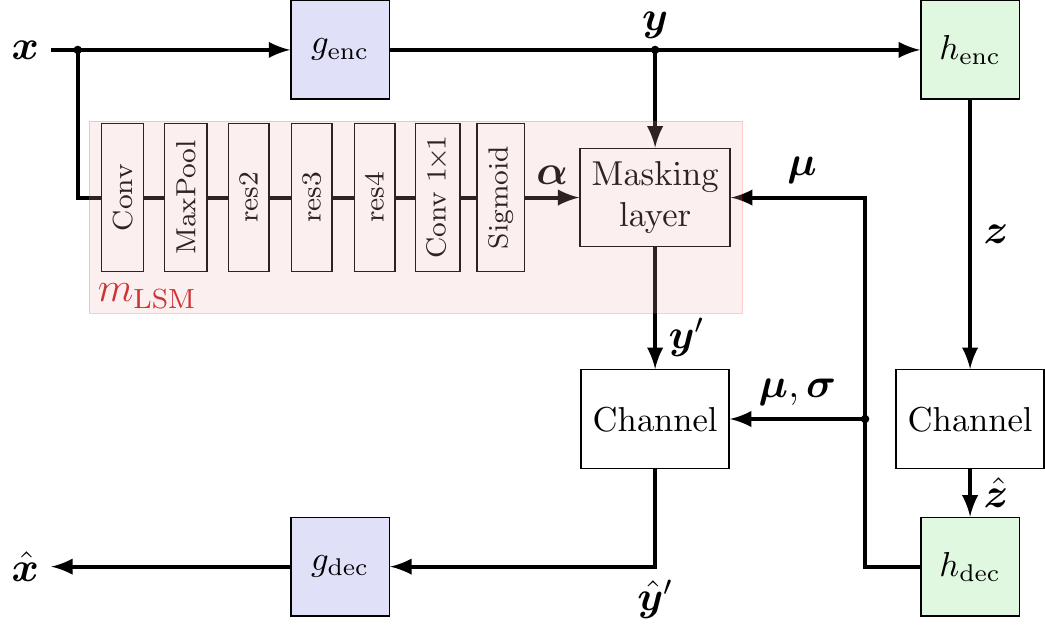}
	\caption{Proposed NCN structure with parallel LSMnet $\parallelNet$ generating masking features~$\mask$ to soft mask the latent representation $\latentRep$ within the range of $\latentRep$ and $\mean$. The channel block comprises the quantization and the arithmetic encoder and decoder.}
	\label{fig:strucure mask LS}
\end{figure}

For the optimization of classic multimedia codecs for VCM, previous approaches~\cite{galteri2018, choi2018, fischer2021_ICASSP} significantly increased the coding gains by reducing the amount of bits that is spent in non-salient areas, which are not important for the analysis network. Inspired by such saliency-driven procedures, we propose LSMnet $\parallelNet$ that generates masking features $\mask$ and runs in parallel to the core NCN encoder $\encoderCoreNcn$. 
%Because of the final sigmoid layer in the parallel branch, $\mask$ is bounded to values between 0 and 1.
The features $\mask$ are obtained to soft mask elements of the latent representation $\latentRep$ that hold information which is presumably not important for the trained analysis task. Thus, such elements of $\latentRep$ can be transmitted with less quality, and therefore less rate, without affecting the detection accuracy of the analysis network. The adapted structure of the NCN with the proposed LSMnet is depicted in Figure~\ref{fig:strucure mask LS}.

\begin{figure}[!t]
	\centering
	\includegraphics[width=0.8\linewidth]{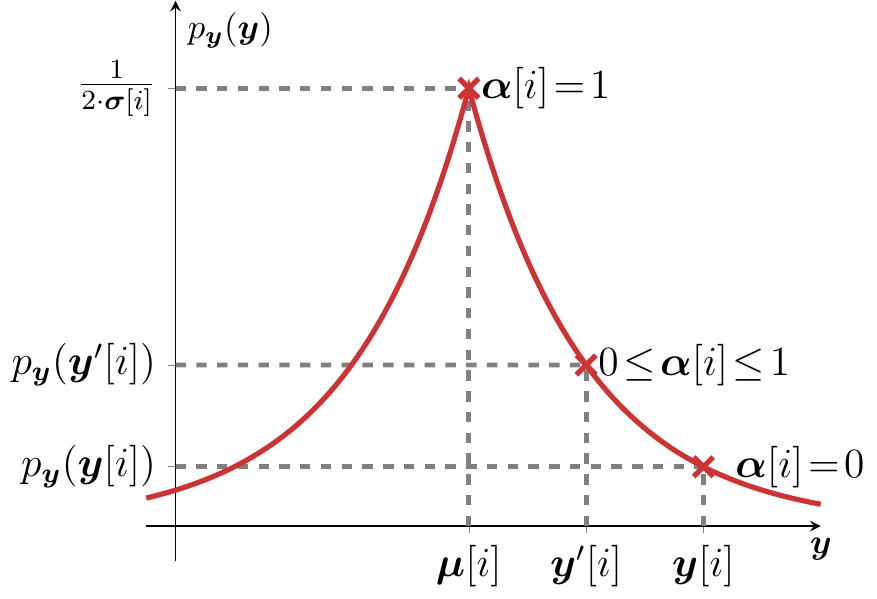}
	\caption{Laplace probability distribution $\probabDist$ for a latent representation $\latentRep$ at position $i$. By soft masking $\latentRep$ with the corresponding mask $\mask[i]$, it can be shifted towards the mean value $\mean[i]$, and thus a higher probability which results in a lower entropy of $\latentRepAdapted[i]$.}
	\label{fig:laplace distribution}
	\vspace{-2mm}
\end{figure}

In the given NCN structure with hyperprior, the amount of bits that has to be spent to transmit one value of the latent representation $\latentRep[i]$ at position $i$ significantly depends on the present probability distribution $\probabDist(\latentRep[i])$ whose main parameters $\mean[i]$ and $\stdDeviation[i]$ are derived from the additionally transmitted hyperprior $\hyperprior$. For our NCN, we employ a Laplace distribution for the latent $\latentRep[i]$ at position $i$ by
\begin{equation}
	\probabDist(\latentRep[i]| \mean[i], \stdDeviation[i]) = \frac{1}{2 \stdDeviation[i]} \cdot  \exp\left( -\frac{\left| \latentRep[i] - \mean[i] \right|}{\stdDeviation[i]}\right),
\end{equation}
which returns a probability for the arithmetic coder to transmit $\latentRep[i]$ to the decoder side. An exemplary distribution for a latent $\latentRep[i]$ is given in Figure~\ref{fig:laplace distribution}. As a rule of thumb it can be assumed that the higher the probability corresponding to the current latent value $\latentRep[i]$, the less bits are required. From this it follows that the more accurate the hyperprior is able to predict the current latent $\latentRep[i]$, the less bitrate is required to transmit $\latentRep[i]$ to the decoder side. From the other perspective of assuming a given distribution and the goal of masking $\latentRep$ such that it requires less bitrate, the non-salient regions cannot simply be zeroed-out, since the Laplace distribution can be of non-zero mean and latents $\latentRep[i]$ far away from the mean $\mean[i]$ result in even more bits to transmit.
% as it can be seen from Figure~\ref{fig:laplace distribution}. 

To overcome this problem, we propose a different soft masking scheme that shifts the latent $\latentRep[i]$ towards the estimated mean value $\mean[i]$ into the adapted value $\latentRepAdapted[i]$ based on the derived masking features $\mask$ by
\begin{equation}
	\latentRepAdapted[i] = \latentRep[i] - \mask[i] \cdot \left( \latentRep[i] - \mean[i] \right).
\end{equation}
With this soft masking scheme as illustrated in Figure~\ref{fig:laplace distribution}, $\latentRep[i]$ either remains untouched for $\mask[i]=0$ and salient areas, or it can be pushed towards its hyperprior estimation $\mean[i]$ for $\mask[i] > 0$. In the extreme case of $\mask[i]=1$, the adapted latent representation $\latentRepAdapted[i]$ equals $\mean[i]$, which means that the adapted latent $\latentRepAdapted[i]$ is entropy coded with the highest probability of $1/(2\cdot\stdDeviation[i])$, resulting in the lowest number of required bits for the given probability distribution. Here, it is important to mention that the decoder does not have to know the mask $\mask$ and thus, no additional information has to be transmitted. The structure of the decoder also remains untouched. In light of a potential deep-learning-based image coding standard, NCNs with parallel LSMnet would still produce standard-compliant bitstreams. 

We propose to derive the masking feature map $\mask$ from the same feature maps that are also used during the inference in the analysis network $\evalNet$. Those features already contain implicit saliency information by focusing on the areas that are important for the evaluation task such as object segmentation. Eventually, an $1\PLH1$ convolution layer and a final Sigmoid layer are applied to those features to generate $\mask$.  With the $1\PLH1$ convolution, the number of channels is reduced to $\numChannels$ and it can be trained to derive which generated features are important for adapting the latent space. Considering the training objective of optimizing the NCN for Mask R-CNN as analysis network, we obtain the first layers of the bottom-up branch of its \hbox{ResNet-50} backbone (ref. Figure~\ref{fig:structure ResNet 50}) to build up the parallel branch $\parallelNet$. We take the features after the \textit{res4} stage, since, they have the same spatial dimensionality $\frac{\height}{16}\PLH\frac{\width}{16}$ as $\latentRep$, which is essential for the masking layer.

One advantage of the proposed masking scheme is that it can be attached retroactively to an already trained NCN. Besides, we only train the weights for the $1\PLH1$ convolution and keep the other weights of $\parallelNet$ fixed, since they already provide sufficient features to derive the saliency information from. With that, the training complexity is kept low. Nevertheless, the $1\PLH1$ convolution can also be trained in conjunction with a fine-tuning of the NCN weights such that the NCN can also adapt towards the additional LSMnet. The inference results of both training methods are presented later in Section~\ref{subsec: Latent Space Masking}. There, we also conduct a cross evaluation to test the influence of different $\parallelNet$ models, which have been trained on different tasks and datasets, on the overall coding performance.

\begin{figure*}[!t]%
	\centering
	\subfloat{\includegraphics[width=0.4\linewidth, valign=c]{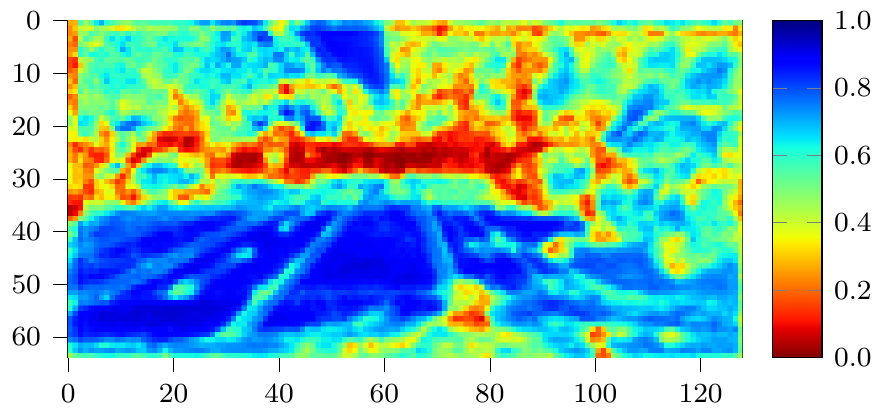} }%
	\,	
	\subfloat{\includegraphics[width=0.28\linewidth, valign=c]{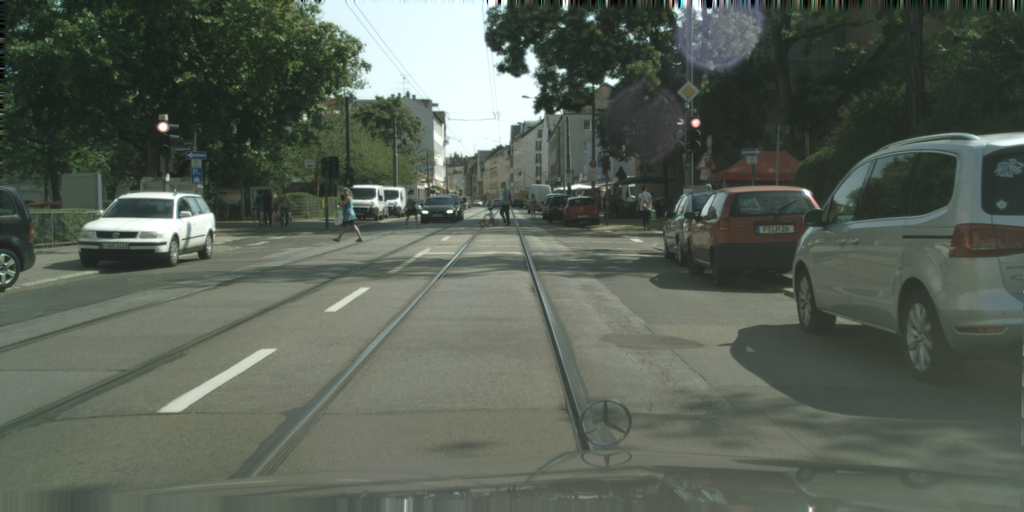} }%
	\,	
	\subfloat{\includegraphics[width=0.28\linewidth, valign=c]{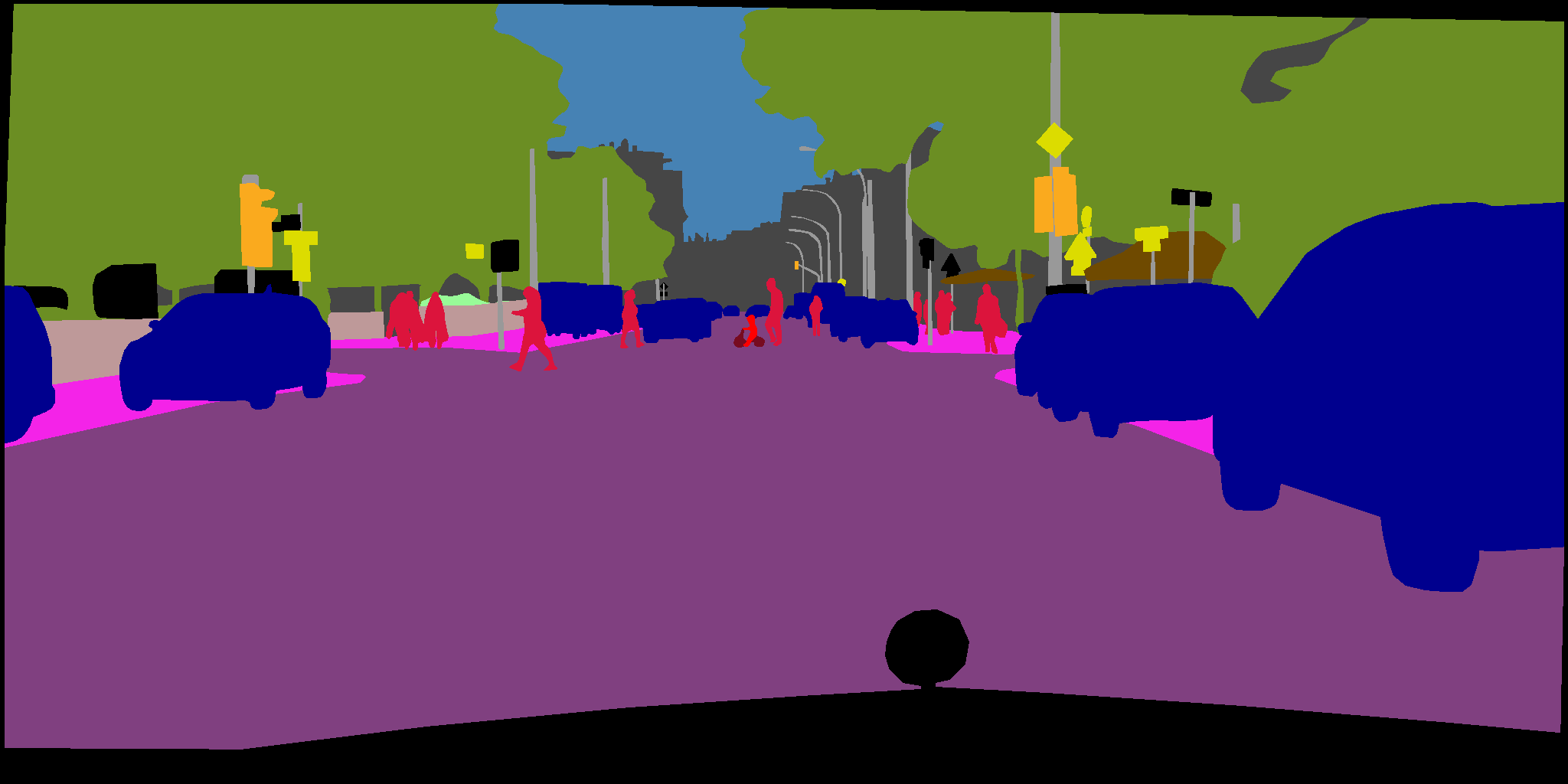} }%
	\caption{Masking features $\mask$ generated by LSMnet averaged over all $\numChannels$ channels (left), for the Cityscapes input image \textit{frankfurt\_000000\_001236\_leftImg8bit} (center). Higher values with blue colors correspond to areas that are considered to be less important by LSMnet. Corresponding ground truth annotations are depicted on the right. \textit{Best to be viewed enlarged on a screen.}}%
	\label{fig:example alpha}%
\end{figure*}

Figure~\ref{fig:example alpha} shows an example of masking features $\mask$ averaged over all $\numChannels$ channels generated by LSMnet. There, the less important areas such as the street or parts of the trees result in a masking value $\mask$ close to one. Furthermore, the additional branch makes a difference between the different sizes of the salient objects. For the larger cars on the right and left side, mostly the edges are considered as salient. But for the small cars and pedestrians in the far horizon of the image, the whole area is considered as salient. This shows that the NCN is able to distinguish from the features generated by the analysis network between areas where the analysis network can easily detect the required object and areas where the analysis network presumably requires a higher quality to get the correct prediction. Ultimately, LSMnet predicts for the first case that a less accurate $\latentRepAdapted$ is sufficient enough for the analysis network to solve the task and thus, the bitrate can be reduced for the corresponding latent.

\section{Training Procedure}
\label{sec:Training Process}
We implemented and trained all our models with the Python deep learning library PyTorch~\cite{paszke2019_PyTorch_short}. % maybe substitute that by footnote
As a basic model, we first trained the NCN as presented in Section~\ref{subsec: Basic Neural Compression Network} for the HVS by minimizing the trainable weights according to~\eqref{eq:loss NCN general} with the distortion from \eqref{eq: loss function HVS} combining MSE and SSIM similar to~\cite{brand2021_CVPR}. The number of channels for the latent representation $N$ was set to 192. As dataset we selected the 800 high-resolution training images from the DIV2K~(D2K) dataset~\cite{agustsson2017_DIV2K} and cropped them to patches of $512\PLH512$ pixels. The batch size was set to four and the training duration was 1000 epochs. All subsequent models were initialized with the weights of this basic model. All models were trained with the Adam optimizer.

In addition, we trained the following three models on the 2965 uncompressed Cityscapes~(CS) training images for 1000 epochs on three different losses:
\begin{itemize}
	\item HVS-based loss (cf. \eqref{eq: loss function HVS}) (NCN+$\lossHvs$+CS)
	\item task-driven loss (cf. \eqref{eq: task loss}) (NCN+$\lossTaskNCN$+CS)
	\item feature-based loss (cf. \eqref{eq: FB loss}) (NCN+$\lossFeature$+CS)
\end{itemize}
For the second model, the pre-trained Mask R-CNN with ResNet-50 backbone and FPN structure from the model zoo of the Detectron2 library\footnote{\url{https://github.com/facebookresearch/detectron2}} was taken to obtain the task loss $\lossTask=\lossMaskRcnn$. The backbone from the same model was taken to generate the \textit{p2} feature space where the SSE was measured to obtain the feature-based loss $\lossFeature$. For the training with Cityscapes data, we randomly cropped the input images to a size of $512\PLH1024$ pixels and also chose a batch size of four. When training the the NCN with \lossTask\ and \lossFeature, no additional regularization with \lossHvs\ has been applied.

To train the proposed additional network LSMnet for latent-space masking, we took the pre-trained models described above and fine-tuned them with the additional branch for another 100 epochs minimizing $\lossTaskNCN$. To initialize LSMnet, we obtained the first layers of already trained out-of-the-shelf models to derive the mask $\mask$ from. In LSMnet, solely the $1\PLH1$ convolution layer is trained in order to keep the complexity low. The weights for the feature generation in LSMnet were taken from the same Detectron2 Mask R-CNN model trained on Cityscapes without further re-training. Besides, we trained another model on the same structure but initialized with weights trained on the COCO~\cite{lin2014_COCO} dataset taken from Detectron2 as well. For the last LSMnet version, we took the features from a VGG-16 network~\cite{simonyan15} trained for the task of image classification on Imagenet~(IN)~\cite{deng2009} from the Torchvision models library\footnote{\url{https://pytorch.org/vision/stable/models.html}}. The three different models were trained with two configurations each. First, only the weights of the $1\PLH1$ convolution were trained. Second, we also allowed for a joint training of the $1\PLH1$ convolution and the NCN weights to analyze whether further coding gains can be achieved by adapting the pre-trained NCN towards the novel encoder structure with LSMnet. 

%Furthermore, we also followed the suggestions in~\cite{le2021_ICASSP} and also trained a model with a task loss that is regularized by a MSE distortion. To that end, we weight the task loss of Mask R-CNN $\lossTask$ and the distortion $\distortion_\mathrm{MSE}$ in the ratio 1:10. We call that loss $\lossTaskMSE$. As a last model, we fine-tuned the model trained on $\lossTaskNCN$ and further fine-tuned the NCN model by also allowing to adapt the weights $\parametersEvalNet$ of the Mask R-CNN analysis model for another 100 epochs. With that, we obtained one general analysis model for all four $\lambda$-values, which is more resilient towards the coding artifacts induced by the NCN compression similar to~\cite{fischer2021_ISCAS}. This model is denoted by the appendix \textit{+ Train Analysis Net}.

\section{Experimental Results}
In the following chapter we present and discuss our experimental results. First, we introduce our experimental setup. Afterwards, we conducted extensive experiments to demonstrate that our proposed methods boost the coding performance for machines. 

\begin{figure*}[!t]%
	\centering
	%	\subfloat[PSNR]{\includegraphics[width=0.33\linewidth, valign=c]{plots/PSNR_over_bitrate.pdf} }%
	%	\subfloat[SSIM]{\includegraphics[width=0.33\linewidth, valign=c]{plots/SSIM_over_bitrate_pdf.pdf} }%
	%%	\subfloat[]{\includegraphics[width=0.33\linewidth, valign=c]{plots/VMAF_over_bitrate.pdf} }%
	%	\subfloat[wAP]{\includegraphics[width=0.33\linewidth, valign=c]{plots/AP_weighted_over_bitrate_avg_pdf.pdf} }%
	%	\quad
	\includegraphics[width=0.99\linewidth]{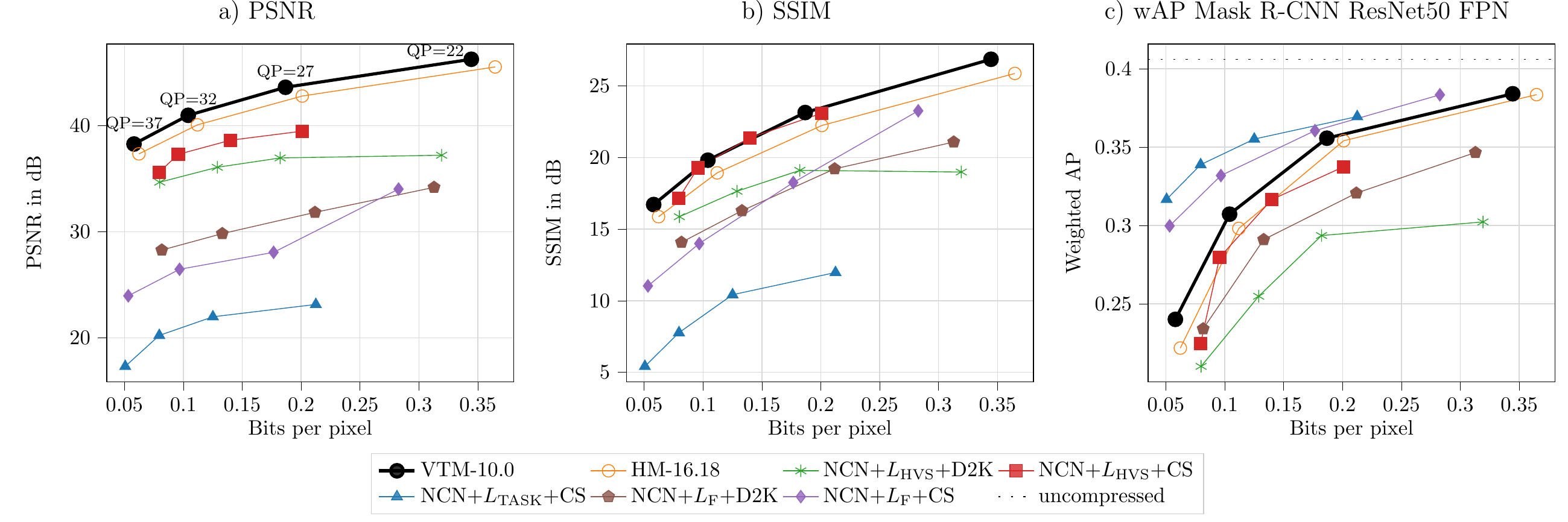}
	%	\vspace{-3mm}
	\caption{Coding results averaged over the 500 Cityscapes validation images for a) PSNR, b) SSIM, and c) wAP as quality metric. The black dotted line in c) denotes the wAP for the Mask R-CNN analysis model on uncompressed images.}%
	\label{fig:plots standard models}%
\end{figure*}

\newcommand{\showmark}[2]{
	\begin{tikzpicture}[baseline, baseline=-.5ex]	
		\draw[#1](0,0) -- (5mm,0); 
		\node[#1, mark size=3, mark options={solid}] at (2.5mm,0){% 
			\pgfuseplotmark{#2}%
		};
	\end{tikzpicture}%
}
\begin{table*}[]
	\centering
	\caption{Bj\o ntegaard delta values for the curves shown in Fig.~\ref{fig:plots standard models} with VTM-10.0 as anchor. BD + metric denotes the difference in the corresponding quality for the same bitrate range. BDR wAP denotes bitrate savings for the same wAP range. The RD-curves for the visual metrics did not allow for a proper BDR measurement. Best values are set in bold.}
	%	\scriptsize
	{\setlength{\extrarowheight}{3pt}% for setting space between the rows
		\begin{tabular}{llll|rrrr|r}
			%	\hline
			\toprule           & \makecell[t]{\\Codec} & \makecell[t]{\\Loss} & \makecell[t]{Training\\Dataset} & \makecell[t]{BD \\ PSNR} & \makecell[t]{BD \\ SSIM} & \makecell[t]{BD \\ VMAF} & \makecell[t]{BD \\ wAP} & \makecell[t]{BDR \\ wAP} \\ \midrule
			\showmark{color1}{o}         & HM-16.18              & -                    & -                               &                 \textbf{-1.2\,dB} &                 -1.3\,dB &                     \textbf{-3.1} &                -1.2\,\% &                 15.3\,\% \\
			\showmark{color2}{asterisk}  & NCN                   & $\lossHvs$           & D2K                             &                 -6.5\,dB &                 -4.2\,dB &                    -18.8 &                -7.0\,\% &                101.5\,\% \\
			\showmark{color3}{square*}   & NCN                   & $\lossHvs$           & CS                              &                 -3.7\,dB &                 \textbf{-0.2\,dB} &                     -8.9 &                -2.1\,\% &                 25.1\,\% \\
			\showmark{color5}{triangle*} & NCN                   & \lossTaskNCN         & CS                              &                -19.9\,dB &                -10.7\,dB &                    -49.4 &                 \textbf{4.0\,\%} &                \textbf{-41.4\,\%} \\
			%	NCN                        & $\lossTaskMSE$        & CS                   & no                              &                -11.1\,dB &                 -8.2\,dB &                    -44.1 &                 4.9\,\% &                -52.5\,\% \\
			%	NCN                        & $\lossTaskNCN$        & CS                   & yes                             &                -21.1\,dB &                -12.1\,dB &                    -47.4 &                 5.2\,\% &                -54.0\,\% \\
			\showmark{color8}{pentagon*} & NCN                   & $\lossFeature$       & D2K                             &                -12.1\,dB &                 -4.8\,dB &                    -26.9 &                -4.1\,\% &                 58.9\,\% \\
			\showmark{color4}{diamond*}  & NCN                   & \lossFeature         & CS                              &                -14.2\,dB &                 -4.8\,dB &                    -31.4 &                 2.4\,\% &                -25.7\,\% \\ \bottomrule
		\end{tabular}
	}
	\label{tab:BD results standard models}
\end{table*}

\subsection{Analytical Methods}
Our NCN models are mainly optimized for the task of instance segmentation with Mask R-CNN. We applied them to the 500 uncompressed Cityscapes validation images in RGB color format. The Cityscapes dataset contains urban street scenes from several German cities with pixel-wise annotated road users allowing to measure the performance for the tasks of instance detection and segmentation. We measured the task performance on the differently compressed input data with the weighted average precision~(wAP) as proposed in~\cite{fischer2020_ICIP}. The wAP is derived from the standard average precision~(AP) that is state of the art to measure the object detection and segmentation performance for well-known datasets such as \commentOut{Pascal VOC~\cite{everingham2014_Pascal_VOC}} Microsoft COCO~\cite{lin2014_COCO}. With the wAP, we weight the AP of each class according to the occurrence frequency of the ground truth instances in order to alleviate class imbalances. For the AP calculation we followed the official Cityscapes code\footnote{\url{https://github.com/mcordts/cityscapesScripts}}. 
%Eventually, the rate-wAP curves are obtained, showing the task performance depending on the required bitrate for each trained network model.

In order to measure the coding performance for the HVS, we select the three commonly used distortion metrics PSNR, SSIM, and VMAF~\cite{li2016_vmaf_short}. All HVS metrics are measured in the YCbCr color space by the VMAF model version 0.6.1\footnote{\url{https://github.com/Netflix/vmaf}}. For all four considered quality metrics PSNR, SSIM, VMAF, and wAP, we measured the Bj\o ntegaard deltas~(BD) in order to quantify the corresponding rate-quality curves. We applied both, the BD-rate~(BDR) measuring the bitrate change for the same quality compared to an anchor, and the BD-quality measuring the quality difference for the same bitrate. Measuring the BD values for AP has been widely used in VCM publications and is also defined in the common testing conditions~(CTCs) of the VCM MPEG group~\cite{liu2020_VCM_CTC}.

As a reference codec we applied the two video coding standards HEVC and VVC with their standard-compliant software implementations HM-16.18\footnote{\url{https://vcgit.hhi.fraunhofer.de/jvet/HM}} and VTM-10.0\footnote{\url{https://vcgit.hhi.fraunhofer.de/jvet/VVCSoftware_VTM}}. Before coding the Cityscapes images, we first transferred them with \textit{ffmpeg} into the YCbCr color space similar to~\cite{fischer2020_ICIP} and as suggested by the MPEG VCM CTCs. Subsequently, the decoded image is transferred back into the RGB color space before applying the analysis network. We encoded each image with the four quantization parameters~(QPs) of 22, 27, 32, and 37 holding on the JVET CTCs. 

When training the NCN models as described in Section~\ref{sec:Training Process}, we trained for four different weighting values $\lambda$. The $\lambda$-values were chosen such that the NCN models result in similar bitrate ranges as the hybrid codecs HM and VTM for the given QPs. Depending on the different losses and training datasets, individual $\lambda$-values had to be chosen for each model.

%% Are BDR/BD_quality values even required for visual metrics????

\begin{figure}[!t]%
	\centering
	\includegraphics[width=0.7\linewidth]{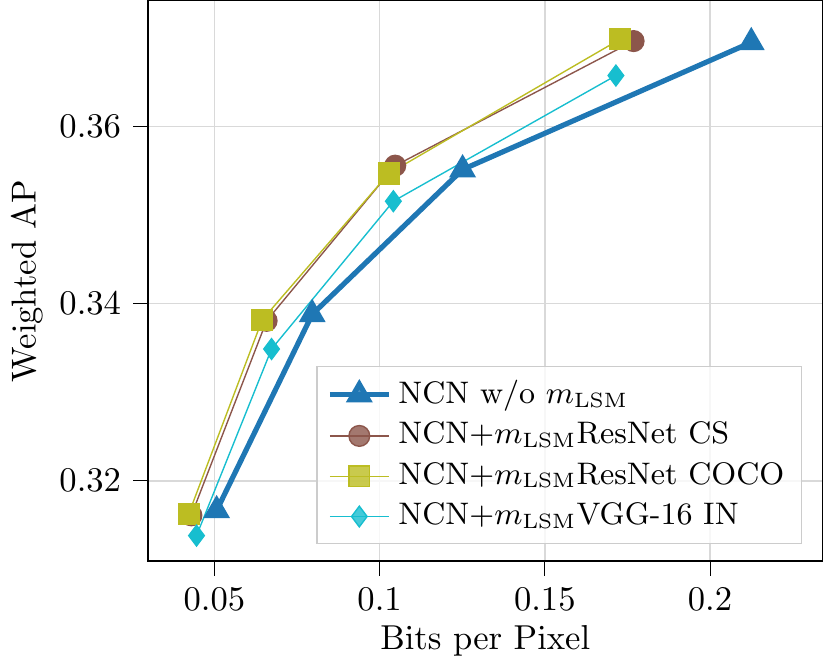}
	\vspace{-3mm}
	\caption{Coding performance for the NCNs with proposed LSMnet with Mask R-CNN ResNet-50 as information sink. NCN without LSMnet is plotted for reference. All models have been trained with $\lossTaskNCN$ and on Cityscapes training data. For NCNs with LSMnet, only the $1\PLH1$ convolution layer has been trained. The weights of the NCN remained frozen for this evaluation.}%
	\label{fig:plots standard mask latent space}%
\end{figure}

\subsection{Adapted Loss Functions}
\label{subsec: Adapted Loss Functions}

First, we evaluate the basic NCN models that have been trained on the three different losses \lossHvs, \lossFeature, and \lossTaskNCN. The results are depicted in Figure~\ref{fig:plots standard models} for PSNR, SSIM, and wAP as quality metric and compared against VTM-10.0 and HM-16.18 as standard hybrid video codec references. The corresponding BD values with the VTM as anchor are listed in Table~\ref{tab:BD results standard models}.  For SSIM and PSNR, the VTM (black) outperforms all NCN variants. Comparing the NCNs, the HVS quality in terms of PSNR and SSIM generated by the models trained on $\lossHvs$ (red and green) highly exceeds the performance of the other NCN models trained for $\lossTaskNCN$ or $\lossFeature$ except for the largest measured bitrate point and SSIM as quality metric. In addition, training the models on the same dataset as for inference, i.e. Cityscapes, also has a large positive impact on the overall coding performance, which proves that initial knowledge about the to-be-expected data characteristics is vital for a high coding efficiency of NCNs. When considering the coding for HVS, the proposed feature-based loss (purple) results in a better visual quality than the task loss (blue). The BD-values for VMAF have the same tendencies.

For VCM, the rate-wAP curves and BD values provide a different picture. When evaluating the coding performance for the Mask R-CNN model as information sink, the NCNs trained with the proposed training with the task-specific (blue) or feature-based losses (purple) outperform the hybrid video codecs. For the standard NCN trained on $\lossTaskNCN$, the wAP is increased by 4.0\,percentage points~(PP) over VTM. For the same NCN trained on the proposed feature-based loss $\lossFeature$, the wAP is enlarged by 2.4\,PP. Reinterpreted, this results in 41.4\,\% and 25.7\,\% rate savings over VTM, respectively. In addition, we also trained a model with the proposed $\lossFeature$ on the DIV2K dataset (brown). 

As the curves in Figure~\ref{fig:plots standard models} prove, training the model with $\lossFeature$ enhances the coding performance over the basic model trained on $\lossHvs$ and DIV2K (green) significantly. This attests that our proposed feature-based loss is indeed able to improve the coding performance compared the normal training with $\lossHvs$ if the evaluation model is already known during training. By doing so, no annotated data is required which is beneficial for practical applications. However, also knowledge about the to-be-expected image characteristics in the form of the chosen training dataset is required for the highest coding gains and to surpass the VTM anchor.
%% Add comments by reviewer 1
There are two possible reasons for the large differences between training the NCN on DIV2K and Cityscapes. First, the NCN model can adapt quite well to the Cityscapes image characteristics during training and thus, eventually perform better during the inference. But, the characteristics and content completely differ between the DIV2K dataset and Cityscapes. Second, since the analysis network is also trained on Cityscapes and its classes, also weaker features for $\lossFeature$ are generated by the analysis network applied to DIV2K data, which contain less objects of the relevant classes. Therefore, the NCN is also trained less effectively on the analysis network and its task.

%Training the NCN on the task-specific loss that is regularized with the MSE (pink) performs even better and results in bitrate savings of 52.5\,\% over VTM indicating that it is indeed beneficial to not completely overfit towards the analysis task. As a last investigation, the NCN model that has been fine-tuned in conjunction with the trainable Mask R-CNN (gray) results in the best coding performance of saving 54\,\% of rate over VTM, when applying the resulting fine-tuned Mask R-CNN model to the deteriorated images.

Comparing those numbers with the already existing end-to-end trained NCN for VCM approaches shows that our model trained on the task loss results in higher BDR savings of 41.4\,\% over VTM-10.0 than the model trained in~\cite{le2021_ICASSP} with 33.7\,\% BDR savings over VTM-8.2. However, the results are hardly comparable, due to the different NCN architectures and experimental setups. When considering the mAP as the authors in~\cite{le2021_ICASSP} instead of the wAP, our bitrate savings marginally change to 42.1\,\%.
The authors of \cite{chamain2021} did not provide any BD results in their work.

\subsection{Latent Space Masking}
\label{subsec: Latent Space Masking}

In this section, we provide measurements to prove that our novel latent space masking network LSMnet boosts the coding performance for machines. All subsequent models have been trained on  $\lossTaskNCN$ and the Cityscapes training data. As a reference model we selected the NCN without LSMnet from the previous analysis (blue curve). Figure~\ref{fig:plots standard mask latent space} shows the corresponding rate-wAP curves, whereas Table~\ref{tab: BD results mask latent space} lists the BD values with the NCN without LSMnet as anchor.

\begin{table}[]
	\centering
	\caption{Bj\o ntegaard delta values for comparing coding performance of NCNs with additional LSMnet. The anchor for the BD measurements is denoted in braces. Compared to Figure~\ref{fig:plots standard mask latent space}, here also the values are shown where the NCN has been fine-tuned in combination with LSMnet. Best values are set in bold.}
	{\setlength{\extrarowheight}{3pt}% for setting space between the rows
		\begin{tabular}{ll|rr|r}
			\toprule
			\makecell[b]{Backbone\\ \parallelNet} & \makecell[b]{Freeze\\ \ncnComplete} & \makecell[b]{BD\\ wAP\\(NCN w/o\\ \parallelNet)} & \makecell[b]{BDR\\ wAP\\(NCN w/o\\ \parallelNet)} & \makecell[b]{BDR\\ wAP\\(VTM-10.0)} \\ \midrule
			ResNet CS                                  & yes                            & 0.7\,\%                    & -16.3\,\%                   & -51.0\,\%                   \\
			ResNet CS                                  & no                           & \textbf{1.1\,\%}                    & \textbf{-27.3\,\%}                   & \textbf{-54.3\,\%}                   \\
			ResNet COCO                                & yes                            & 0.7\,\%                    & -17.4\,\%                   & -51.6\,\%                   \\
			ResNet COCO                                & no                           & 1.0\,\%                    & -25.5\,\%                   & \textbf{-54.3\,\%}                   \\
			VGG-16 IN                                  & yes                            & 0.3\,\%                    & -7.1\,\%                    & -47.8\,\%                   \\
			VGG-16 IN                                  & no                           & -0.1\,\%                   & 3.7\,\%                     & -40.6\,\%                   \\
			\bottomrule
		\end{tabular}
	}
	\label{tab: BD results mask latent space}
\end{table}

All curves in Figure~\ref{fig:plots standard mask latent space} show the performances when only the $1\PLH1$ convolution layer has been trained during the additional 100 epochs of training. In general, all three NCNs with parallel LSMnet outperform the reference NCN without LSMnet. Comparing the same measurement points generated by the different $\lambda$-values during training with and without LSMnet shows that the bitrate is substantially reduced by masking out the non-salient latents, while the detection accuracy of the Mask R-CNN is still preserved. Thereby, the amount of required bitrate is similar for all three types of LSMnet. However, the resulting wAP is lower for the LSMnet, where $\mask$ is derived from the first features of the VGG-16 network that has been trained for image classification on Imagenet. Nevertheless, this results in 7.1\,\% bitrate savings compared to the reference NCN without LSMnet. This underlines that even CNNs that have not been trained for the exact inference task can be taken for LSMnet to derive the saliency information from. When deriving the features from the same ResNet backbone as it is used in the analysis Mask R-CNN, higher BDR savings can be obtained. The results indicate that it does not make a large difference if the LSMnet backbone has been trained on the exact same dataset and classes as used in the inference. All in all, 16.3\,\% of bitrate can be saved by only training the $1\PLH1$ convolution layer of LSMnet.

For a further analysis, we also allow the training optimizer to adapt the NCN weights together with LSMnet. As the BD values in Table~\ref{tab: BD results mask latent space} demonstrate, this results in even higher coding gains of up to 27.3\,\%. All in all, taking the standard VTM as reference, we obtain up to 54.3\,\% of bitrate savings by our NCN architecture with the novel LSMnet deriving the saliency masks from the same backbone as used in inference.

\subsection{Cross Evaluation}

With the previous investigations, we have shown that our proposed methods strongly boost the coding performance of NCNs for machines. There, the analysis model that has been employed during training to adapt the loss functions for the VCM task remained the same for the inference. Here, we now analyze how well the different NCN models, which are all strongly optimized for the Detectron2 Mask R-CNN \hbox{ResNet-50 FPN}, perform for other analysis models on the two VCM evaluation tasks of instance segmentation and object detection.

\setcounter{table}{2}
\begin{table*}[]
	\centering
	\caption{BDR values for NCN cross evaluation for several analysis networks and VTM-10.0 as anchor. All models have been trained on Cityscapes and Mask R-CNN ResNet-50 FPN as analysis network. V-39 denotes the VoVNet-39 backbone. For the NCN with LSMnet, the ResNet trained on CS was taken and fine-tuning of the NCN was allowed. Best values are set in bold.}
	{\setlength{\extrarowheight}{3pt}% for setting space between the rows
		\begin{tabular}{ll|r|rrrrr}
			\toprule
			Loss     & LSMnet & \makecell[b]{Mask R-CNN \\ResNet-50 FPN} & \makecell[b]{Mask R-CNN \\ V-39 FPN} & \makecell[b]{Faster R-CNN \\ResNet-50 FPN} & \makecell[b]{Faster R-CNN \\V-39 FPN} & \makecell[b]{CenterMask \\ResNet-50 FPN} & \makecell[b]{CenterMask \\V-39 FPN} \commentOut{& \makecell[b]{DeepLab V3+\\ResNet-103} }\\ \midrule
			$\lossHvs$      & no     & 25.1\,\%                    & \textbf{35.7\,\%}                 & 37.7\,\%                      & \textbf{43.2\,\%}                   & 25.0\,\%                    & \textbf{28.5\,\%}                 \\
			$\lossTaskNCN$     & no     & -41.4\,\%                   & 133.4\,\%                & -37.8\,\%                    & 274.9\,\%                  & -37.6\,\%                   & 79.0\,\%       \\
			%		TASK+MSE & no     & -52.5\,\%                   & -8.05\,\%                & -53.1\,\%                     & -9.0\,\%                   & -49.4\,\%                   & XX\,\%                   & 15.9\,\%        \\
			$\lossTaskNCN$     & yes    & \textbf{-54.3\,\%}                   & 80.0\,\%                 & \textbf{-56.7\,\%}                     & 188.5\,\%                  & \textbf{-52.9\,\%}                   &  31.7\,\%       \\     
			$\lossFeature$       & no     & -25.7\,\%                   & 64.1\,\%                 & -29.9\,\%                     & 79.2\,\%                   & -32.6\,\%                   &  45.6\,\%          \\ \bottomrule
			
		\end{tabular}
	}
	\label{tab: BD results for cross evaluation}
\end{table*}

For the first task, Mask R-CNN and CenterMask~\cite{lee2020_CenterMask} were obtained. Faster R-CNN~\cite{ren2017} was employed for object detection. For all three networks, we trained a model with a ResNet-50 FPN and a VoVNet-39 FPN~\cite{lee2019_VoVNet} backbone each. We trained all five new models on the Cityscapes dataset with the Detectron2 library and its suggested configurations.

The BDR results with VTM-10.0 as anchor are listed in Table~\ref{tab: BD results for cross evaluation}. The coding gains of the differently trained NCN models measured for Faster R-CNN and CenterMask are pretty similar to the coding gains measured for the optimized Mask R-CNN analysis network as long as the same ResNet-50 is taken as backbone. When taking the analysis networks with VoVNet-39 as backbone, the coding performance of the NCNs significantly drops compared to the networks with ResNet-50 backbone. These results indicate that the end-to-end optimization on the Mask R-CNN with the task-specific loss model primary results in output images from which the ResNet-50 FPN backbone can derive optimal features from. Therefore, all investigated network architectures can still perform their task from the resulting features as long as the NCNs are trained on the same backbone network as used in inference. This is the case even though the Mask R-CNN, Faster R-CNN, and CenterMask models share the same backbone structure but not the exact weights. These observations hold for both VCM optimization losses of $\lossTaskNCN$ and $\lossFeature$, with $\lossFeature$ resulting in less coding deficits since it is not as specialized as a training with the task-driven loss. The results indicate that the VCM-specialized training, regardless whether with $\lossTaskNCN$ or $\lossFeature$ mainly results in compressed images that are strongly adapted to the ResNet-50 backbone such that it can still derive high-quality features from the deteriorated input images. Employing other backbone structures at the decoder side results in worse coding performance, since the characteristics of the decoded images do not fit to other backbones than the trained one. 

\subsection{Visual Examples}

\def\imageSize{0.315}
\setcounter{table}{1}
\begin{figure*}
	%\begin{table*}
	\begin{tabular}{ccc}
		\centering
		\includegraphics[width=\imageSize\linewidth]{visual_results/orig.png} & 
		\includegraphics[width=\imageSize\linewidth]{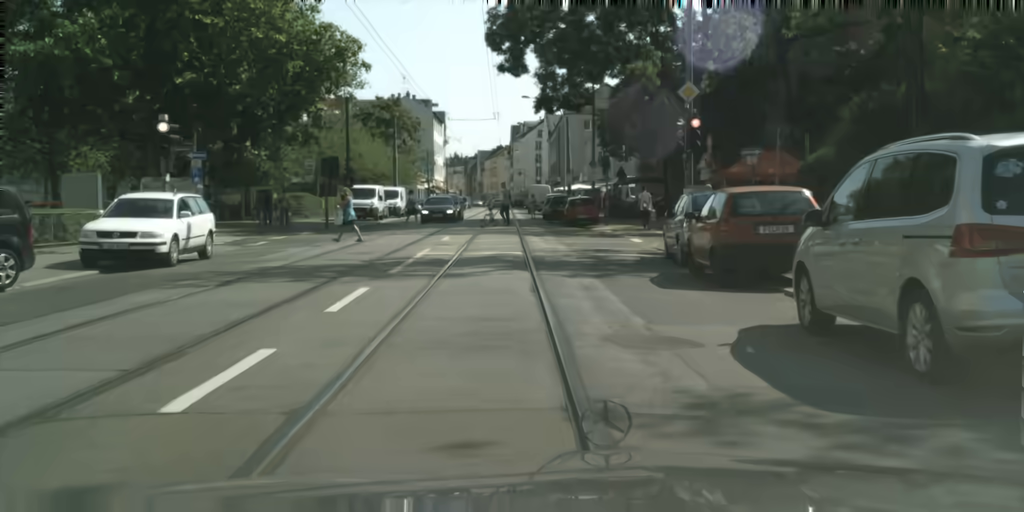} & 
		\includegraphics[width=\imageSize\linewidth]{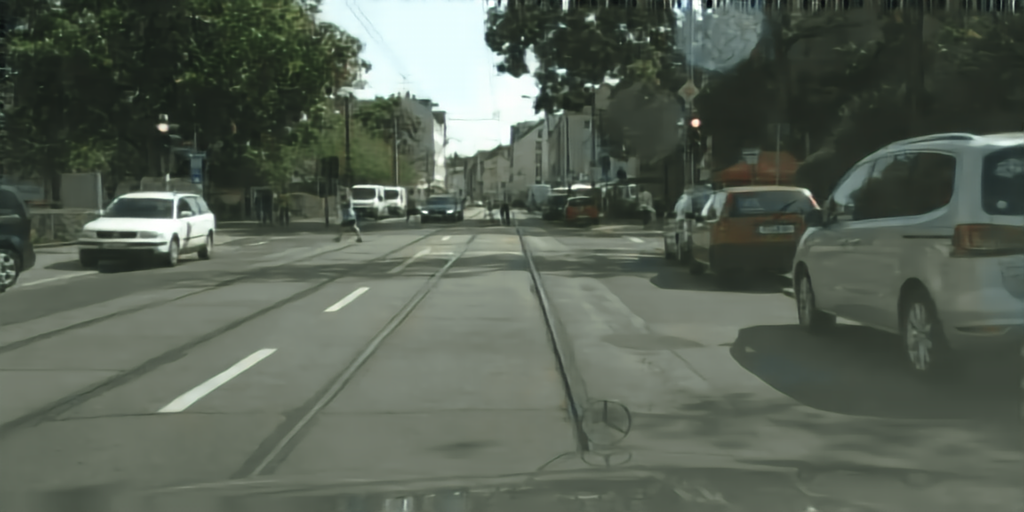} \\
		a) Original image & b) VTM-10.0 (QP=37) @0.057\,bpp & c) NCN+\lossHvs+CS @0.045\,bpp \medskip\\
		
		\includegraphics[width=\imageSize\linewidth]{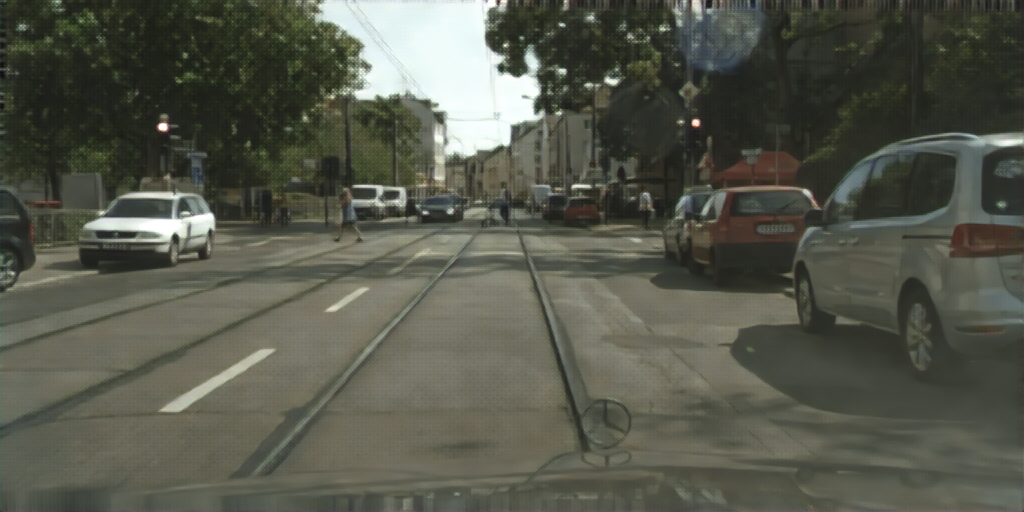} & 
		\includegraphics[width=\imageSize\linewidth]{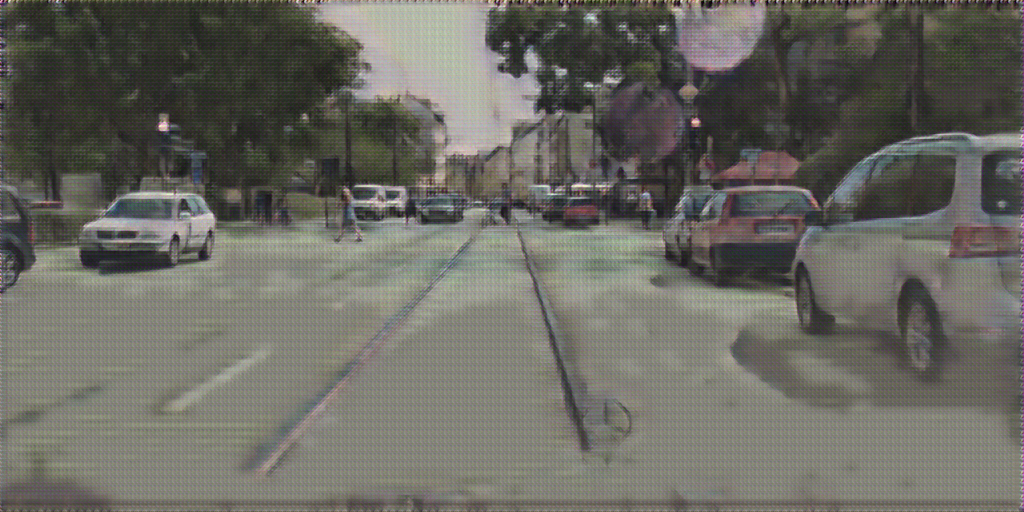} & 
		\includegraphics[width=\imageSize\linewidth]{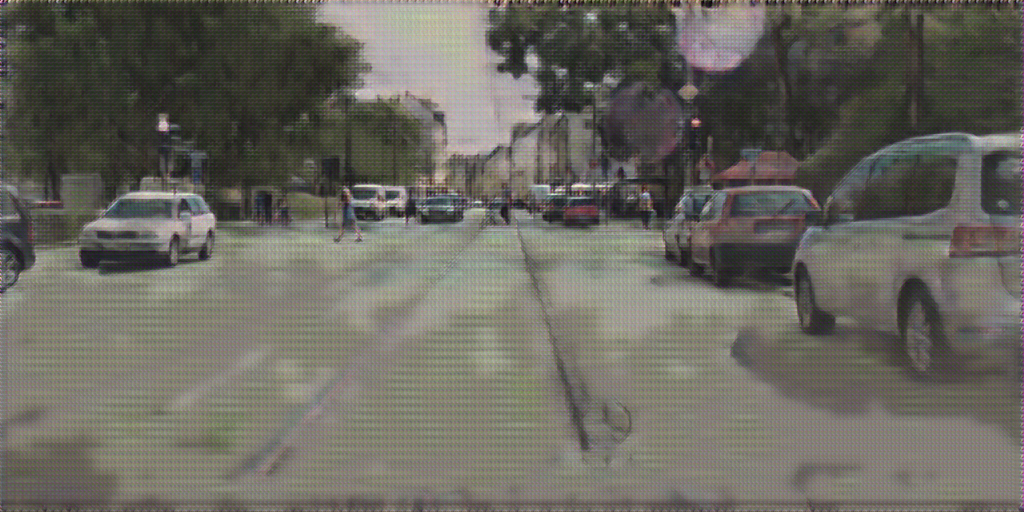} \\
		d) NCN+\lossFeature+CS @0.052\,bpp & e) NCN+\lossTaskNCN+CS @0.045\,bpp & f) NCN+\lossTaskNCN+LSMnet+CS @0.039\,bpp \medskip\\
		
		\includegraphics[width=\imageSize\linewidth]{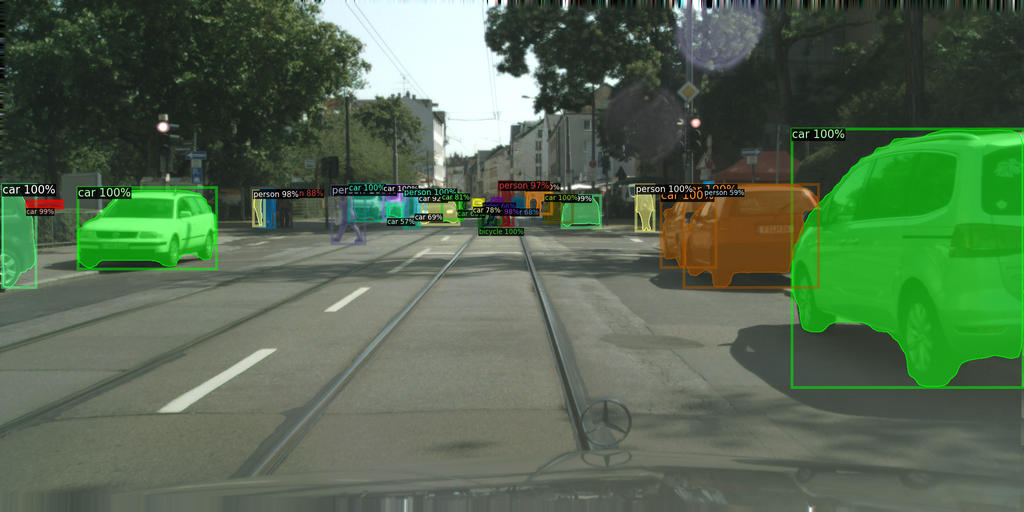} & 
		\includegraphics[width=\imageSize\linewidth]{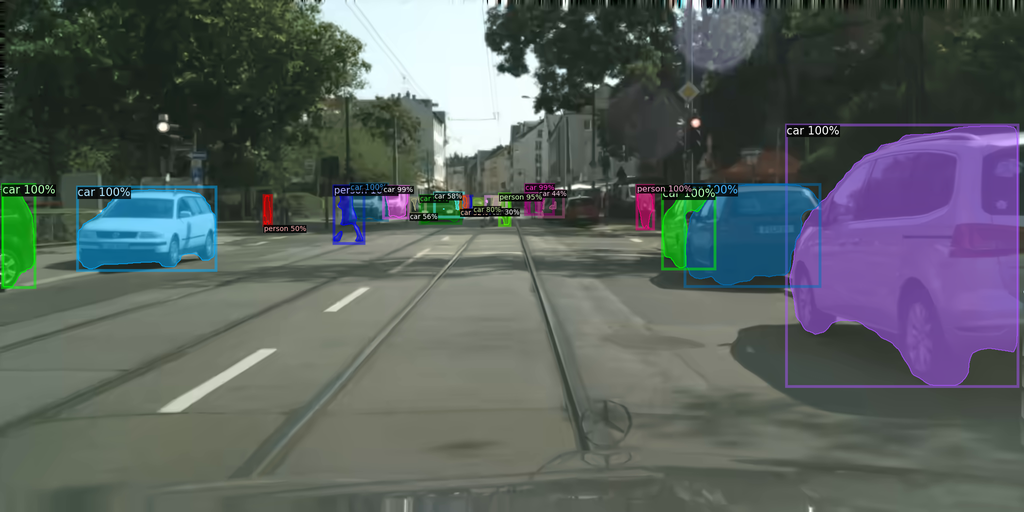} & 
		\includegraphics[width=\imageSize\linewidth]{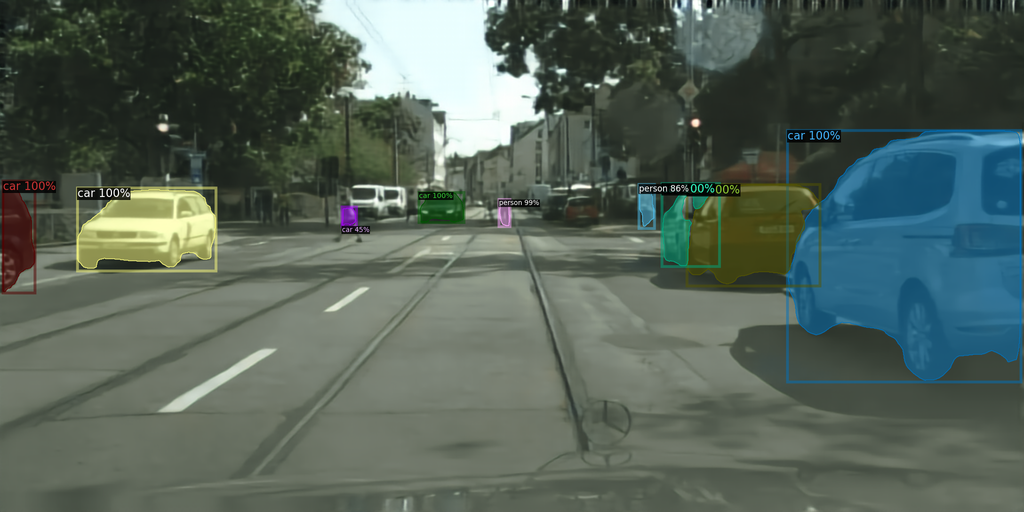}\\
		g) Detections original image & h) Detections VTM-10.0 (QP=37) & i) Detections NCN+\lossHvs+CS \\
		24\,TP + 3\,FP + 8\,FN & 13\,TP + 4\,FP + 19\,FN @0.057\,bpp & 7\,TP + 2\,FP + 25\,FN @0.045\,bpp\medskip\\
		
		\includegraphics[width=\imageSize\linewidth]{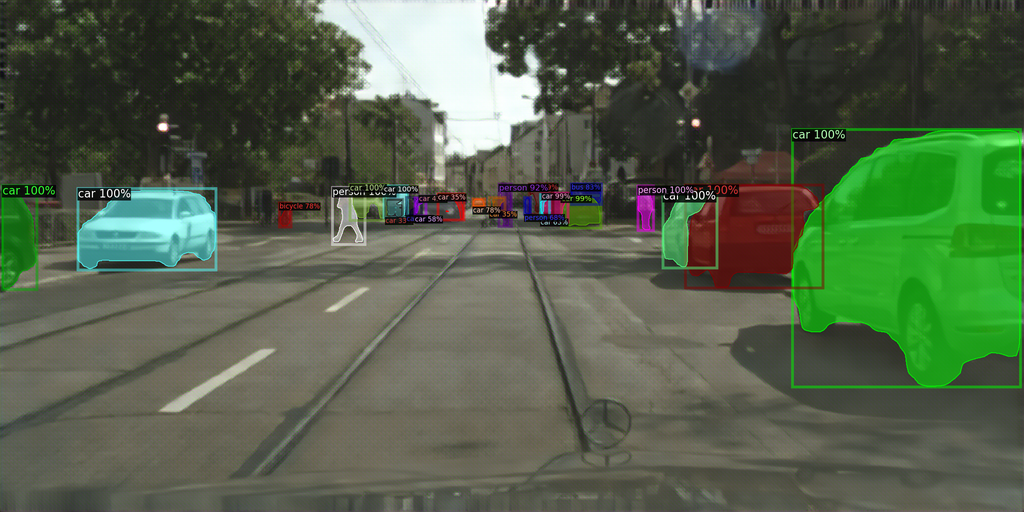} & 
		\includegraphics[width=\imageSize\linewidth]{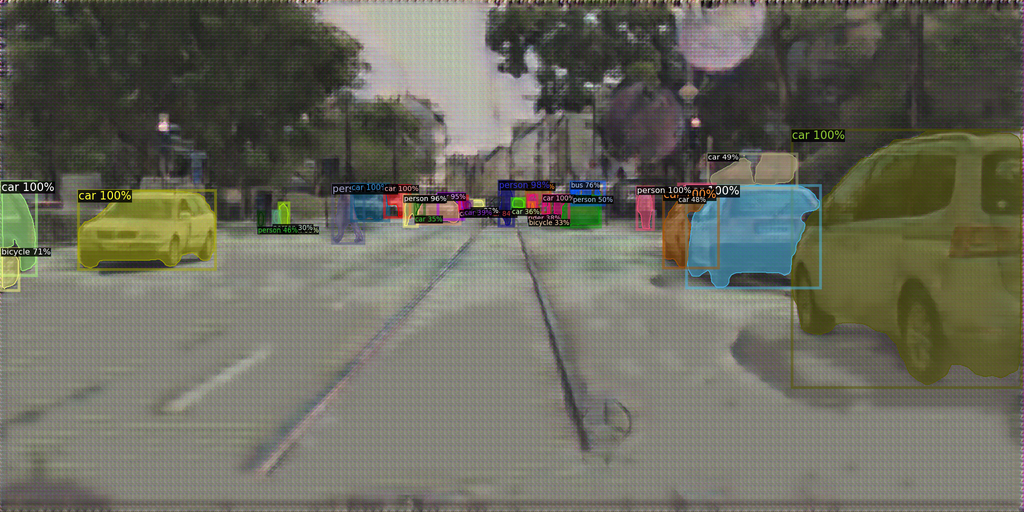} & 
		\includegraphics[width=\imageSize\linewidth]{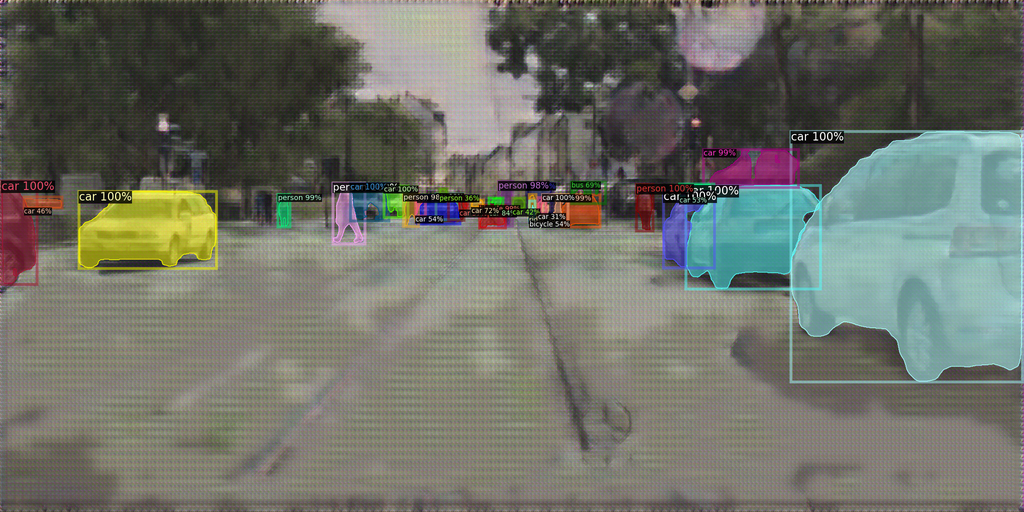} \\
		j) Detections NCN+\lossFeature+CS & k) Detections NCN+\lossTaskNCN+CS & l) Detections NCN+\lossTaskNCN+LSMnet+CS \\
		16\,TP + 6\,FP + 16\,FN @0.052\,bpp & 18\,TP + 8\,FP + 14\,FN @0.045\,bpp & 18\,TP + 9\,FP + 14\,FN @0.039\,bpp \\ 					
	\end{tabular}
	\caption{Visual results for the Cityscapes image \textit{frankfurt\_000000\_001236\_leftImg8bit}. The images a) to f) show the decoded output images generated by the codecs. The images g) to l) show the detections of Mask R-CNN with ResNet-50 FPN applied to the corresponding decoded images. bpp denotes bits per pixel. For the NCN with LSMnet, fine-tuning of the NCN was allowed. The image has 32 annotated ground truth instances: 12 persons, 1 rider, 18 cars, and 1 bicycle. 
		Below each image with detections the true positive~(TP), false positive~(FP), and false negatives~(FN) are given for an IoU threshold of 0.5.
		\textit{Best to be viewed enlarged on a screen.}}
	\label{fig:visual results}
	%\end{table*}
\end{figure*}

Figure~\ref{fig:visual results} gives a visual impression for an exemplary Cityscapes image that is coded with VTM-10.0 and four differently trained NCNs at low bitrate. In the lower part, the corresponding Mask R-CNN predictions are shown.

The images show that important objects such as cars and pedestrians appear to have sharper edges in the images that are coded by the NCNs trained for VCM compared to the VTM or the NCN trained for the HVS loss. Besides the color values of those objects seem over-saturated. To compensate for that, non-salient areas such as trees with a high structure requiring more bitrate are coded with less visual quality. Moreover, a strong overlaying pattern can be recognized by the VCM-optimized NCNs, which is possibly generated to result in high-fidelity features in the initial ResNet-50 layers of the analysis network. This supports the explanations from the previous cross evaluation section on the weaker coding performance of the NCNs for other inference analysis networks with different backbones as listed in Table~\ref{tab: BD results for cross evaluation}. Due to the different structure of a VoVNet-39 backbone, another pattern would have to be created, to obtain high fidelity features, and therewith a high detection accuracy. Furthermore, comparing the NCN with and without LSMnet shows that non-salient areas such as the road marking or the rails are coded with less quality, when adding LSMnet to the NCN structure. This results in less bitrate by maintaining a similar detection accuracy of Mask R-CNN.

Despite the strong overfitting on the VCM task, all coded images in Figure~\ref{fig:visual results} could also still be analyzed by e.g. a human supervisor or a court of law to comprehend the actions of the machine. Therefore, the proposed neural compression framework for machines can be interpreted as an intermediate step between the two coding paradigms of \textit{analyze-then-compress} and \textit{compress-then-analyze} explained in the beginning. The large coding gains for the VCM scenarios we have achieved throughout the paper are based on a strongly adapted representation for an analysis network with a certain backbone, which is very closely related to the feature compression. Nevertheless, this strongly adapted representation is restricted to an image in the RGB color space and could still be interpreted by other network structures or humans. However, less task accuracy or visual quality is achieved for those cases. In general, this is not possible for standard feature coding, since the sender has to transmit the exact features that the receiver is able to interpret.

%%%%% Example Figure %%%%%%%%
%\begin{figure}[!t]
%\centering
%\includegraphics[width=2.5in]{fig1}
%\caption{Simulation results for the network.}
%\label{fig_1}
%\end{figure}

%%%%%%%%%%% Example Figure over two cols %%%%%%%%%%%%%%%%%
%\begin{figure*}[!t]
%\centering
%\subfloat[]{\includegraphics[width=2.5in]{fig1}%
%\label{fig_first_case}}
%\hfil
%\subfloat[]{\includegraphics[width=2.5in]{fig1}%
%\label{fig_second_case}}
%\caption{Dae. Ad quatur autat ut porepel itemoles dolor autem fuga. Bus quia con nessunti as remo di quatus non perum que nimus. (a) Case I. (b) Case II.}
%\label{fig_sim}
%\end{figure*}

\section{Conclusion}
In this paper, we have shown several methods to enhance the compression performance of NCNs when coding for a neural network as information sink by sacrificing visual quality. To that end, we employed an end-to-end training procedure allowing to optimize the NCN for a Mask R-CNN applied to the decoded output image. For that, the inference evaluation network has to be known before the training, which is usually given in practical scenarios. This end-to-end VCM training resulted in coding gains of 41.4\,\% compared to standard \hbox{VTM-10.0}. In addition, we also proposed a novel feature-based loss that allows for a VCM-optimized training without requiring annotated training data but also saved 25.7\,\% bitrate. As our main contribution, we proposed the parallel masking network LSMnet that allows for a soft masking of the latent space in order to incorporate saliency information within the coding process for an excelled coding performance. To derive the masking features, we employed the first layers of CNNs that have been trained for the tasks of image classification or instance segmentation since they already implicitly generate saliency information representing the later analysis task at the decoder side. Due to its modular structure, LSMnet can easily be added to further NCN structures without requiring  a complete new training of the NCN model. Compared to the NCN without LSMnet, additional 27.3\,\% were saved resulting in bitrate savings of 54.3\,\% against \hbox{VTM-10.0}. For even larger coding gains, LSMnet could be upgraded by utilizing attention models to add more global information to the masking features. Besides, we strongly believe that LSMnet can also be applied in non-VCM scenarios. Future research will show whether the general idea of soft masking the latent space can also be deployed for human visual coding. Nevertheless, optimal losses have to be found to support this approach.

\section*{Acknowledgments}
The authors gratefully acknowledge that this work has been supported by the Deutsche Forschungsgemeinschaft (DFG, German Research Foundation) under project number 426084215.

\bibliographystyle{IEEEtran}

\begin{IEEEbiography}[{\includegraphics[width=1in,height=1.25in,clip,keepaspectratio]{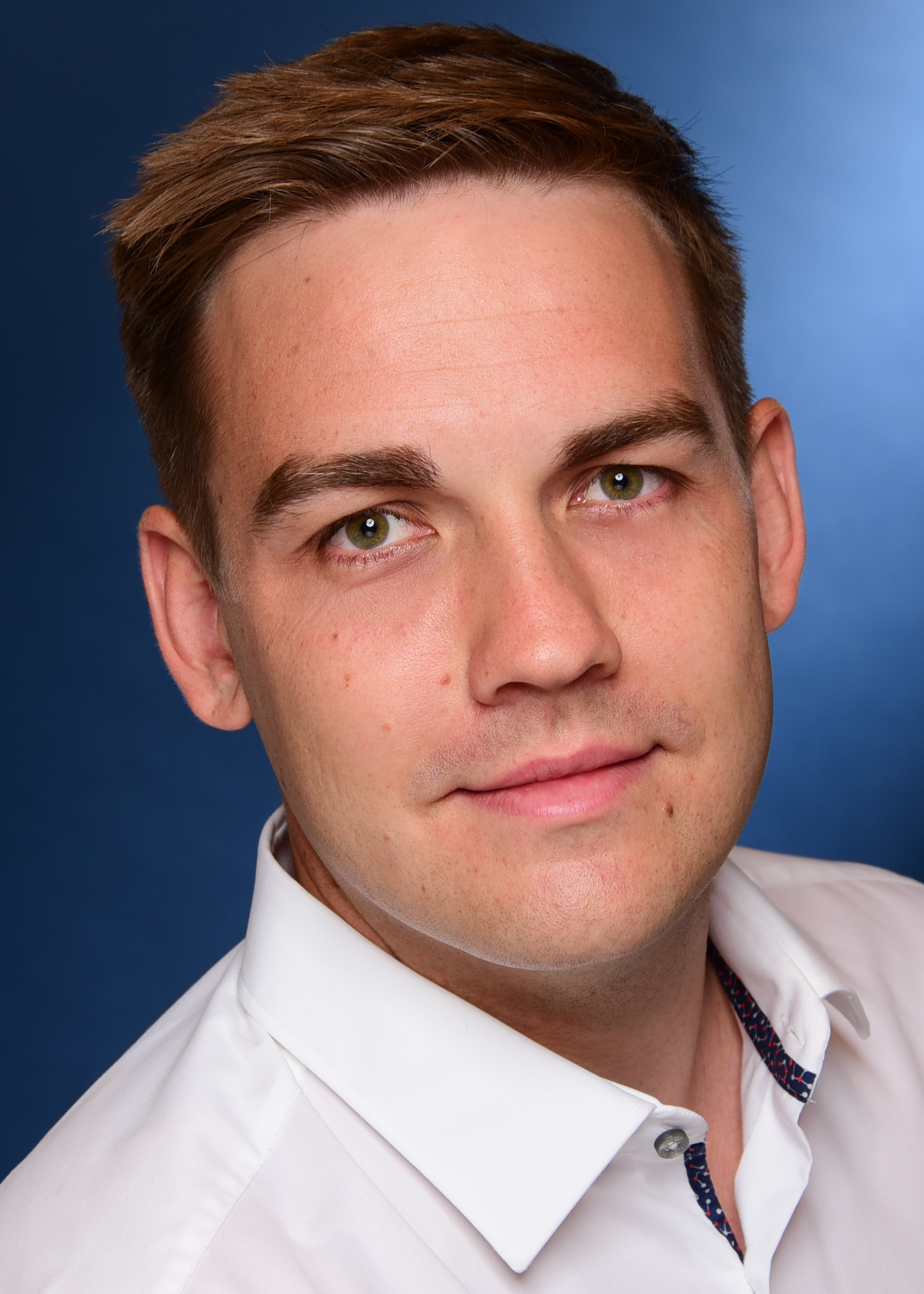}}]{Kristian Fischer}
	(Graduate Student Member, IEEE) received the B.Eng degree in electrical engineering and information technology from University of Applied Sciences in Aschaffenburg and the M.Sc degree in electrical engineering from Friedrich-Alexander-University Erlangen-N\"urnberg (FAU), in 2015 and 2017, respectively. His studies also included a semester abroad at the Minnesota State University Mankato in the USA in 2016. Since 2018, he has been a research scientist with the Chair of Multimedia Communications and Signal Processing, Friedrich-Alexander-University Erlangen-N\"urnberg. His research interests include image and video compression for machines and deep neural networks in particular.
	
\end{IEEEbiography}

\begin{IEEEbiography}[{\includegraphics[width=1in,height=1.25in,clip,keepaspectratio]{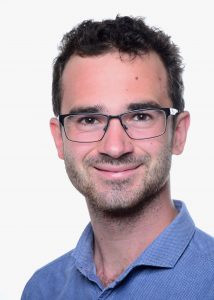}}]{Fabian Brand}
	(Graduate Student Member, IEEE) received his master's degree in electrical engineering from Friedrich-Alexander University Erlangen-N\"{u}rnberg  (FAU), Germany, in 2018. During his bachelors, he worked on methods for frame-rate-conversion of video sequences and during his masters he researched automated harmonic analysis of classical music and style classification. Since 2019, he  has  been  a  researcher  with  the  Chair  of Multimedia Communications and Signal Processing, Friedrich-Alexander University Erlangen-N\"{u}rnberg (FAU), where he conducts research on methods for video compression and deep learning. For his work, among others, he received the best paper award of the Picture Coding Symposium (PCS) 2019.
	
\end{IEEEbiography}

\begin{IEEEbiography}[{\includegraphics[width=1in,height=1.25in,clip,keepaspectratio]{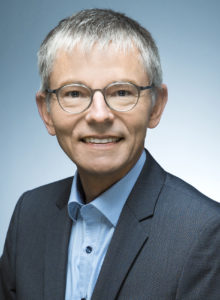}}]{Andr\'e Kaup}
	(Fellow, IEEE) received the Dipl.-Ing. and Dr.-Ing. degrees in electrical engineering from RWTH Aachen University, Aachen, Germany, in 1989 and 1995, respectively.
	
	He joined Siemens Corporate Technology, Munich, Germany, in 1995, and became the Head of the Mobile Applications and Services Group in 1999. Since 2001, he has been a Full Professor and the Head of the Chair of Multimedia Communications and Signal Processing at Friedrich-Alexander University Erlangen-Nuremberg (FAU), Germany. From 2005 to 2007 he was Vice Speaker of the DFG Collaborative Research Center 603. From 2015 to 2017, he served as the Head of the Department of Electrical Engineering and Vice Dean of the Faculty of Engineering at FAU.
	
	Dr. Kaup is a member of the Scientific Advisory Board of the German VDE/ITG and he served as a member of the IEEE Multimedia Signal Processing Technical Committee. He is a member of the editorial board of the IEEE Circuits and System Magazine and he served as an Associate Editor of the IEEE Transactions on Circuits and Systems for Video Technology. He was a Guest Editor for IEEE Journal of Selected Topics in Signal Processing. He was a Siemens Inventor of the Year 1998 and obtained the 1999 ITG Award. He received several IEEE Best Paper Awards including the Paul Dan Cristea Special Award in 2013, and his group won the Grand Video Compression Challenge from the Picture Coding Symposium 2013. The Faculty of Engineering with FAU and the State of Bavaria honored him with Teaching Awards, in 2015 and 2020, respectively. He is an IEEE Fellow and a member of the Bavarian Academy of Sciences.
	
	He has authored around 400 journal and conference papers and has over 120 patents granted or pending. His research interests include image and video signal processing and coding, and multimedia communication.
\end{IEEEbiography} 

\vfill

\end{document}